\documentclass[useAMS,usenatbib,twocolappendix]{emulateapj}

\usepackage{ulem}
\usepackage{color}
\usepackage{graphics,graphicx}
\usepackage{times} 
\usepackage{amssymb}
\usepackage{amsmath}
\usepackage{natbib}
\usepackage{url}
\usepackage{multirow}
\usepackage{ctable}
\usepackage{threeparttable}
\newif\ifAMStwofonts
\AMStwofontstrue

\def\xmm{{\it XMM-Newton}}

\def\suzaku{{\it Suzaku}}

\def\epicpn{{EPIC-pn}}
\def\epicmos1{{EPIC-MOS1}}
\def\epicmos2{{EPIC-MOS2}}
\def\epicmos{{EPIC-MOS}}

\def\nustar{{\it NuSTAR}}

\def\m{{$\rm\thinspace m$}}

\def\pcmsq{\hbox{$\rm\thinspace cm^{-2}$}}
\def\H0{{\rm ~km~s^{-1}~Mpc^{-1}}}

\def\kev{\hbox{\rm keV}}

\def\ergpcmsqps{\hbox{erg~cm$^{-2}$~s$^{-1}$}}

\def\ergps{\hbox{erg~s$^{-1}$}}

\def\msun{\hbox{$M_{\odot}$}}

\def\chisq{{$\chi^{2}$}}

\def\nustardas{\rm {\small NUSTARDAS}}

\def\flx2xsp{\rm{\small FLX2XSP}}

\def\presto{\rm {\small PRESTO}}
\def\accelsearch{\rm {\small ACCELSEARCH}}

\def\grid25{\hbox{\rm{\small GRID25}}}

\def\simpl{\rm{\small SIMPL}}

\def\tbabs{\rm{\small TBABS}}
\def\tbnew{\rm {\small TBNEW}}

\def\bb{\rm{\small BB}}
\def\diskbb{\rm{\small DISKBB}}
\def\diskpbb{\rm{\small DISKPBB}}

\def\cutoffpl{\rm{\small CUTOFFPL}}

\def\heii{\rm He\,{\small II}}

\def\eg{{\it e.g.}}

\def\ie{{\it i.e.~\/}}

\def\la{\mathrel{\hbox{\rlap{\hbox{\lower4pt\hbox{$\sim$}}}{\raise2pt\hbox{$<$}}}}}
\def\ga{\mathrel{\hbox{\rlap{\hbox{\lower4pt\hbox{$\sim$}}}{\raise2pt\hbox{$>$}}}}}

\def\d25{D$_{25}$}

\def\.25{0.25 keV\thinspace}

\def\rsp{$R_{\rm{sp}}$}
\def\rmag{$R_{\rm{M}}$}
\def\rco{$R_{\rm{co}}$}

\def\ngc{NGC\,5907 ULX}
\def\p13{NGC\,7793 P13}
\def\m82{M82 X-2}

\shorttitle{Evidence for Pulsar-like Emission Components in the Broadband ULX Sample}
\shortauthors{D.~J. Walton et al.}

\begin{document}

\title{Evidence for Pulsar-like Emission Components in the Broadband ULX Sample}

\author{D. J. Walton\altaffilmark{1},
F. F\"urst\altaffilmark{2},
M. Heida\altaffilmark{3},
F. A. Harrison\altaffilmark{3},
D. Barret\altaffilmark{4,5},
D. Stern\altaffilmark{6},
M. Bachetti\altaffilmark{7},
M. Brightman\altaffilmark{3},\\
A. C. Fabian\altaffilmark{1},
M. J. Middleton\altaffilmark{8}
}
\affil{ \\
$^{1}$ Institute of Astronomy, University of Cambridge, Madingley Road, Cambridge CB3 0HA, UK \\
$^{2}$ European Space Astronomy Centre (ESA/ESAC), Operations Department, Villanueva de la Ca\~nada (Madrid), Spain \\
$^{3}$ Space Radiation Laboratory, California Institute of Technology, Pasadena, CA 91125, USA \\
$^{4}$ Universite de Toulouse; UPS-OMP; IRAP; Toulouse, France \\
$^{5}$ CNRS; IRAP; 9 Av. colonel Roche, BP 44346, F-31028 Toulouse cedex 4, France \\
$^{6}$ Jet Propulsion Laboratory, California Institute of Technology, Pasadena, CA 91109, USA \\
$^{7}$ INAF/Osservatorio Astronomico di Cagliari, via della Scienza 5, I-09047 Selargius (CA), Italy \\
$^{8}$ Department of Physics and Astronomy, University of Southampton, Highfield, Southampton SO17 1BJ, UK \\
}

\begin{abstract}
We present broadband X-ray analyses of a sample of bright ultraluminous X-ray
sources with the goal of investigating the spectral similarity of this population to the
known ULX pulsars, \m82, \p13\ and \ngc. We perform a phase-resolved analysis of
the broadband \xmm+\nustar\ dataset of \ngc, finding that the pulsed emission from
the accretion column in this source exhibits a similar spectral shape to that seen in
both \m82\ and \p13, and that this is responsible for the excess emission observed at
the highest energies when the spectra are fit with accretion disk models.
We then demonstrate that similar `hard' excesses are seen in all the ULXs in the
broadband sample. Finally, for the ULXs where the nature of the accretor is currently
unknown, we test whether the hard excesses are all consistent with being produced
by an accretion column similar to those present in \m82, \p13\ and \ngc. Based on the
average shape of the pulsed emission, we find that in all cases a similar accretion
column can successfully reproduce the observed data, consistent with the hypothesis
that this ULX sample may be dominated by neutron star accretors. Compared to the
known pulsar ULXs, our spectral fits for the remaining ULXs suggest that the
non-pulsed emission from the accretion flow beyond the magnetosphere makes a
stronger relative contribution than the component associated with the accretion
column. If these sources do also contain neutron star accretors, this may help to
explain the lack of detected pulsations.
\end{abstract}


\section{Introduction}

The discovery that three ultraluminous X-ray sources (ULXs) are powered by
accreting pulsars -- \m82: \citealt{Bachetti14nat}, \p13: \citealt{Fuerst16p13,
Israel17p13}, and \ngc: \citealt{Israel17} -- has brought about a paradigm shift in our
understanding of this exotic population. ULXs appear to radiate in excess of the
Eddington limit for the standard $\sim$10\,\msun\ stellar remnant black holes seen
in Galactic X-ray binaries (\ie $L_{\rm{X}} > 10^{39}$\,\ergps), and so black hole
accretors had generally been assumed. The brightest ULXs were previously
considered to be good candidates for intermediate mass black holes ($10^2 \lesssim
M_{\rm{BH}} \lesssim 10^5$\,\msun; \citealt{Sutton12}) based on their extreme
luminosities. However, the pulsar \ngc\ has an apparent peak X-ray luminosity of
$\sim$7 $\times 10^{40}$\,\ergps\ (assuming isotropy; \citealt{Israel17,
Fuerst17ngc5907}), making it one of the brightest ULXs known (\eg\ \citealt{Swartz04,
WaltonULXCat}). These objects are therefore extreme, with luminosities up to
$\sim$500 times the Eddington limit for a standard 1.4\,\msun\ neutron star
($\sim$2$\times$10$^{38}$\,\ergps).

Given that all three sources were known to be ULXs long before their identification as
pulsars (\eg\ \citealt{Kaaret09m82, Motch14nat, Sutton13}), it is natural to ask how
many other members of the ULX population could also be powered by accretion onto
a neutron star. Although pulsations have not currently been detected from any other
members of the ULX population (\citealt{Doroshenko15}), the pulsations in both \m82\
and \ngc\ are transient (\citealt{Bachetti14nat, Israel17}). Exactly why this is the case is
not currently well understood. Nevertheless, a lack of observed pulsations therefore
does not exclude a neutron star accretor. \cite{King17ulx} suggest that even for ULXs
that are powered by neutron stars, pulsations may only be observable when the
magnetospheric radius is close to (or larger than) the spherization radius. The former
is the point at which the accretion disk is truncated by the magnetic field of the neutron
star, and the material is forced to follow the field lines instead, and the latter is the point
at which the accretion disk transitions to the thick inner flow expected for
super-Eddington accretion (\eg\ \citealt{Shakura73, Abram88, Dotan11}). This idea
appears to be supported by our spectral analysis of \p13\ (\citealt{Walton18p13}).

Since it may not always be possible to identify neutron star ULXs through the
detection of pulsations, and dynamical mass measurements are challenging owing to
the faint stellar counterparts (\eg\ \citealt{Gladstone13, Heida14, Lopez17}), other
methods for identifying neutron star ULXs will be of key importance for our
understanding of the nature of this population as a whole. Here, we investigate
potential spectral signatures of neutron star ULXs by determining the properties of the
pulsed emission from the three known systems, and assessing whether similar
features are seen in the broader ULX population for which the nature of the accretors
remains unknown. The paper is structured as follows: in Section \ref{sec_5907} we
perform a phase-resolved analysis of the ULX pulsar \ngc\ for comparison with \m82\
and \p13, in Section \ref{sec_sample_hardex} we compare the high-energy properties
of the known ULX pulsars and the general ULX population, and in Section
\ref{sec_sample_ULXp} we present fits to the broadband ULX sample with models
directly motivated by the known ULX pulsars. We discuss our results in Section
\ref{sec_dis} and summarise our conclusions in Section \ref{sec_conc}.

\section{The Pulsed Emission from NGC\,5907 ULX}
\label{sec_5907}

We begin by investigating the pulsed emission from \ngc\ in order to separate the
emission from the accretion column (pulsed) from the rest of the accretion flow
(assumed to be steady over the pulse cycle). For this analysis, we focus on the
high-flux coordinated observation performed with the \nustar\ (\citealt{NUSTAR}) and
\xmm\ (\citealt{XMM}) observatories in 2014 (a log of all the observations considered
in this work is given in Appendix \ref{app_obslog}), as this is the only broadband
observation of \ngc\ in which the pulsations have been detected to date. We refer the
reader to \cite{Fuerst17ngc5907} for details on the reduction of these data, but note
that here we are only able to utilize data from the \epicpn\ detector (\citealt{XMM_PN})
for the lower energy \xmm\ data, since the pulse period is shorter than the time
resolution of the \epicmos\ detectors.

\subsection{Difference Spectroscopy}
\label{sec_pulse_diff}

First, we isolate the spectrum of the pulsed component, similar to our recent analyses
of \m82\ and \p13\ (\citealt{Brightman16m82a, Walton18p13}). To do so, we follow the
same approach as in those works and extract spectra from the brightest and the
faintest quarters ($\Delta\phi_{\rm{pulse}} = 0.25$) of the pulse cycle (the pulse-profile
of \ngc\ is nearly sinusoidal, similar to \p13\ and \m82; see Figure 1 in
\citealt{Israel17}), and subtract the latter from the former (\ie ``pulse-on"$-$``pulse-off").
Given the low signal-to-noise, we combine the data from the \nustar\ FPMA and FPMB
detectors for \ngc\ throughout this work. The combined data are rebinned to have a
minimum signal-to-noise (S/N) per energy bin of 3 to ensure the errors are at
least close to Gaussian and allow the use of \chisq\ statistics. We fit the data over the
$\sim$0.5--25\,keV energy range with a simple \cutoffpl\ model, allowing for neutral
absorption from the Galactic column, fixed to $N_{\rm{H, Gal}} = 1.2 \times
10^{20}$\,\pcmsq\ (\citealt{NH}), and intrinsic to NGC\,5907 ($N_{\rm{H, int}}$; $z = 
0.002225$) using the \tbnew\ absorption code. We adopt the abundances of
\cite{tbabs} and the cross-sections of \cite{Verner96} throughout this work. However,
the intrinsic column is poorly constrained in these fits, so we fix it to $7 \times
10^{21}$\,\pcmsq\ following previous analyses of the average broadband spectra
(\citealt{Walton15, Fuerst17ngc5907}).

\begin{figure}
\begin{center}
\hspace*{-0.6cm}
\rotatebox{0}{
{\includegraphics[width=235pt]{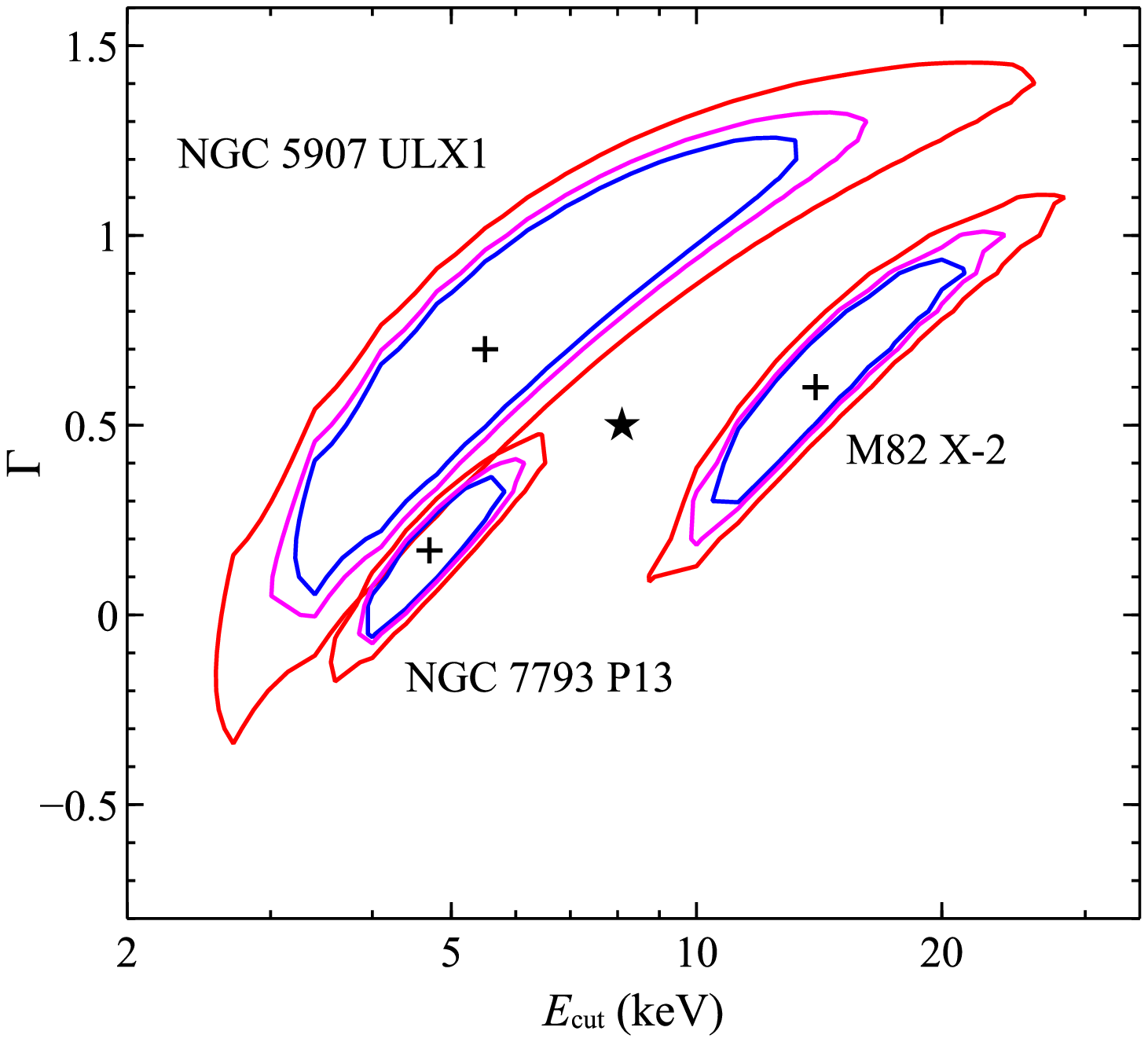}}
}
\end{center}
\caption{
2D confidence contours for $\Gamma$ and $E_{\rm{cut}}$ for the pulsed
spectra of M82\,X-2, NGC\,7793 P13 and NGC\,5907 ULX1. The 90, 95 and 99\%
confidence contours for 2 parameters of interest are shown in blue, magenta and
red, respectively. The star shows the average of the best-fit values, which is utilized
in our fits to the broadband ULX sample (Section \ref{sec_sample_ULXp}).}
\vspace{0.3cm}
\label{fig_pulse_gamEc}
\end{figure}

As with both \m82\ and \p13, the simple \cutoffpl\ model provides a good fit to the \ngc\
pulsed spectrum, with $\chi^{2} = 24$ for 21 degrees of freedom (DoF). Although the
constraints are not as strong in this case, the results are also broadly similar to both
\m82\ and \p13; the pulsed spectrum shows a hard rise before exhibiting a cutoff at
fairly low energies: $\Gamma = 0.7^{+0.4}_{-0.5}$, $E_{\rm{cut}} =
5.5^{+4.8}_{-2.1}$\,keV. We show the 2-D confidence contours for $\Gamma$ and
$E_{\rm{cut}}$ in Figure \ref{fig_pulse_gamEc} for all three of the pulsar ULXs currently
known. While there are quantitative differences between them, all three sit in broadly
the same area of parameter space in terms of their pulsed emission. For further
comparison, the pulsed flux from \ngc\ during this epoch corresponds to an apparent
0.5--25.0\,keV luminosity of $2.6^{+0.3}_{-0.4} \times 10^{40}$\,\ergps\ (assuming
isotropy) for a distance to NGC\,5907 of 17.1\,Mpc (\citealt{Tully16}). This is significantly
more luminous than the pulsed emission in both \m82\ and \p13, although this is not
surprising given that \ngc\ is inferred to have a much higher phase-averaged luminosity
than these other systems.

\begin{table*}
  \caption{Best fit parameters obtained from our phase-resolved analysis of the broadband 2014 observation of \ngc}
  \vspace{-0.3cm}
\begin{center}
\begin{tabular}{c c c c c c c c c c}
\hline
\hline
\\[-0.1cm]
Model & \multicolumn{2}{c}{Parameter} & \multicolumn{2}{c}{Stable Continuum:} \\
\\[-0.2cm]
Component & & & \diskpbb\ & \diskbb+BB \\
\\[-0.2cm]
\hline
\hline
\\[-0.1cm]
\tbabs\ & $N_{\rm H; int}$ & [$10^{21}$ cm$^{-2}$] & $6.7^{+0.7}_{-0.5}$ & $7.5^{+1.4}_{-0.8}$ \\
\\[-0.1cm]
\diskpbb/\diskbb\ & $kT_{\rm{in}}$ & [keV] & $2.8^{+0.5}_{-0.8}$ & $0.4 \pm 0.1$ \\
\\[-0.1cm]
& $p$ & & $0.57^{+0.04}_{-0.07}$ & 0.75 (fixed) \\
\\[-0.1cm]
& Norm & & $3.2^{+2.5}_{-1.9} \times 10^{-3}$ & $0.8^{+3.4}_{-0.5}$ \\
\\[-0.1cm]
BB & $kT$ & [keV] & -- & $1.2 \pm 0.2$ \\
\\[-0.1cm]
& Norm & [$10^{-6}$] & -- & $4.5 \pm 1.3$ \\
\\[-0.1cm]
\cutoffpl\ & $\Gamma$ & & \multicolumn{2}{c}{$0.7$\tmark[a]} \\
\\[-0.1cm]
& $E_{\rm{cut}}$ & [keV] & \multicolumn{2}{c}{$5.5$\tmark[a]} \\
\\[-0.1cm]
& $F_{2-10}$\tmark[b] (low) & [$10^{-12}$\,\ergpcmsqps] & $<1.1$ & $1.0 \pm 0.1$ \\
\\[-0.1cm]
& $F_{2-10}$\tmark[b] (med) & [$10^{-12}$\,\ergpcmsqps] & $1.1^{+0.3}_{-0.8}$ & $1.3 \pm 0.1$ \\
\\[-0.1cm]
& $F_{2-10}$\tmark[b] (high) & [$10^{-12}$\,\ergpcmsqps] & $1.3^{+0.4}_{-0.8}$ & $1.6 \pm 0.1$ \\
\\[-0.2cm]
\hline
\\[-0.1cm]
$\chi^{2}$/DoF & & & 383/385 & 370/384 \\
\\[-0.2cm]
\hline
\hline
\\[-0.4cm]
\end{tabular}
\label{tab_5907_phaseres}
\end{center}
\flushleft
$^a$ These parameters have been fixed to the best-fit values from the ``pulse
on"$-$``pulse-off" difference spectroscopy (Section \ref{sec_pulse_diff}). \\
$^b$ Observed fluxes for the CUTOFFPL component in the 2-10\,keV band.
\vspace{0.3cm}
\end{table*}

\subsection{Phase-Resolved Spectroscopy}
\label{sec_5907_phaseres}

We also perform phase-resolved spectroscopy of \ngc, again following our analysis of
\p13. We continue using phase bins of $\Delta\phi_{\rm{pulse}} = 0.25$ in size, and
extract spectra from three different fluxes across the pulse cycle: high-flux (cycle peak),
medium-flux (rise $+$ fall) and low-flux (cycle minimum).\footnote{As with \p13, we
initially extracted the `rise' and `fall' spectra separately, but on inspection their spectra
were found to be similar, and so were combined into a single medium-flux dataset.} We
fit the broadband spectra from all three phase bins simultaneously and undertake a
simple decomposition of the data into stable (non-pulsed) and variable (pulsed)
components. Based on our ``pulse on"$-$``pulse-off" difference spectroscopy (Section
\ref{sec_pulse_diff}), we treat the pulsed emission from the accretion column with a
\cutoffpl\ model. The `shape' parameters for this component ($\Gamma$, $E_{\rm{cut}}$)
are linked across all three phase bins and are fixed to the results obtained above. The
normalisation of the \cutoffpl\ component can vary between the phase bins, and we do
not require this to be zero for the low-flux data, as the emission from the accretion
column could still contribute during the minimum of the pulse cycle. 

For the stable emission, which we expect to come from the accretion flow outside of
\rmag, we initially fit a single \diskpbb\ component, following the time-averaged
analyses in \cite{Walton15} and \cite{Fuerst17ngc5907}. This is a multi-color accretion
disk model in which the radial temperature index of the disk ($p$) is free to vary in
addition to its inner temperature ($T_{\rm{in}}$) and normalisation. All the parameters
for this component are linked across all three phase bins, as is the intrinsic neutral
absorption column. The global fit to the phase-resolved data with this model is very
good, $\chi^{2}$/DoF = 383/385, and the parameter results are presented in Table
\ref{tab_5907_phaseres}. For the \diskpbb\ component, we find that the inner
temperature and radial temperature index for this component are similar to those
presented in \cite{Fuerst17ngc5907}: $T_{\rm{in}} = 2.8^{+0.5}_{-0.8}$\,keV and $p =
0.57^{+0.04}_{-0.07}$. The latter is flatter than expected for a thin disk (which should
give $p = 0.75$; \citealt{Shakura73}), which could imply the presence of a thick,
advection dominated accretion disk, as expected for super-Eddington accretion (\eg\
\citealt{Abram88}). There is a strong degeneracy between the stable \diskpbb\
component and the zero-point of the variations from the pulsed \cutoffpl\ component,
resulting in large errors on their respective normalisations/fluxes, as for the parameter
combination found in the ``pulse on"$-$``pulse-off" spectroscopy the \diskpbb\ model
can produce a similar shape to the \cutoffpl\ model. However, for the best-fit
parameters, we find that the accretion column (the \cutoffpl\ component) dominates
the flux at the highest energies probed by \nustar, exactly where the high-energy
powerlaw tail reported by \cite{Fuerst17ngc5907} contributes in their model for the
phase-averaged data.

\begin{figure}
\begin{center}
\hspace*{-0.5cm}
\rotatebox{0}{
{\includegraphics[width=240pt]{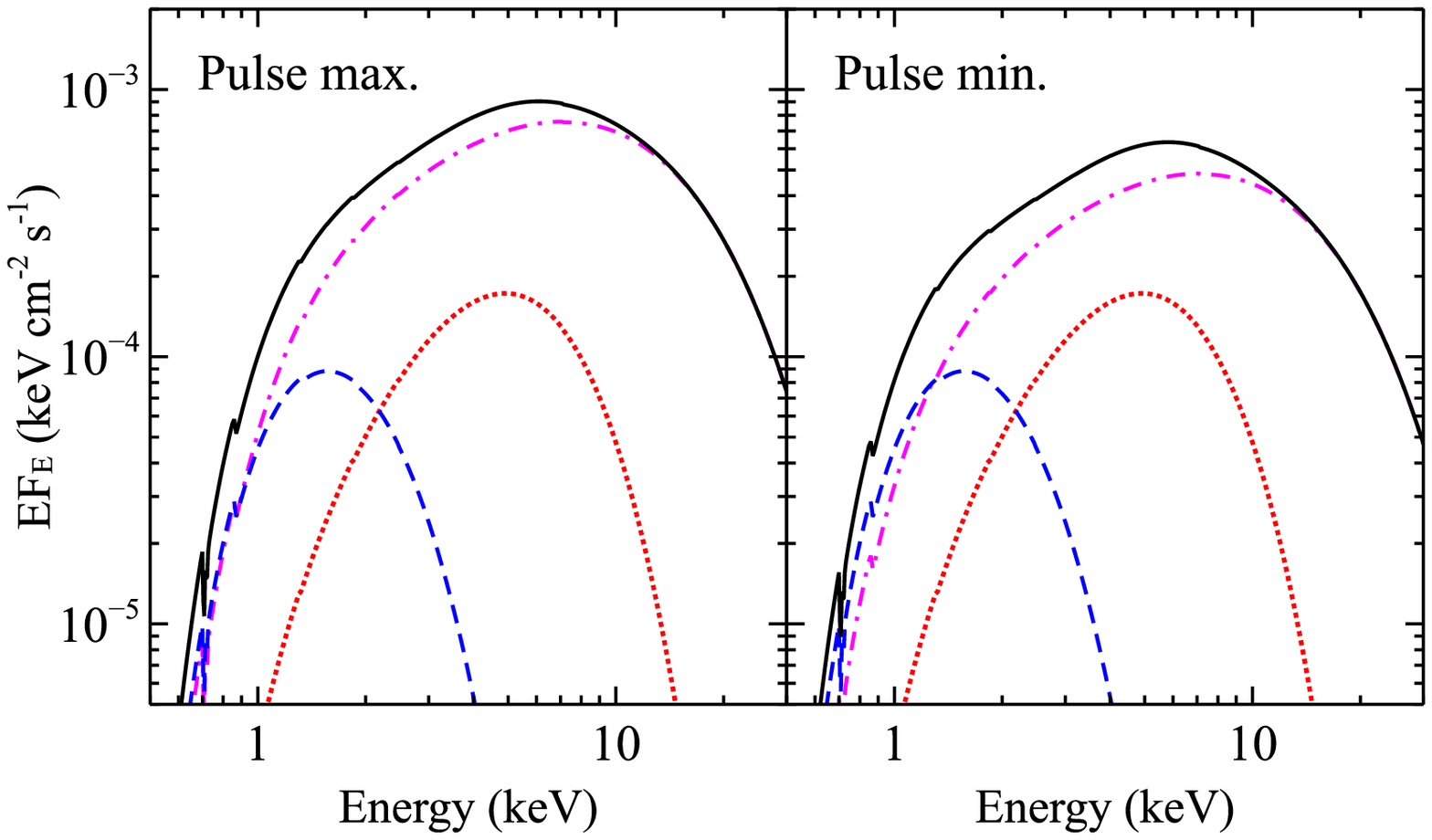}}
}
\end{center}
\caption{
The relative contributions of the various spectral components during the peaks
(\textit{left panel}) and the troughs (\textit{right panel}) of the pulse cycle from our
phase-resolved analysis of \ngc\ with the DISKBB+BB+CUTOFFPL continuum
model (similar to Figure 4 in \citealt{Walton18p13} for \p13). In both panels the total
model is shown in solid black, the DISKBB and BB components (both steady) are
shown in dashed blue and dotted red, respectively, and the CUTOFFPL component
(pulsed) in dash-dot magenta. As the DISKBB and DISKPBB components are assumed 
to be steady across the pulse cycle, these components are identical in both panels.}
\label{fig_5907_phaseres}
\vspace{0.3cm}
\end{figure}

Motivated by our results for \p13, which require two thermal blackbody components to
fit the observed spectra (\citealt{Walton18p13}), we also fit a second model in which
the stable emission for \ngc\ consists of two thermal components: a thin disk (\diskbb)
and a hotter blackbody (BB) component. As before, all the parameters for these
components are the same for each phase bin. The global fit is also excellent with this
model, $\chi^{2}$/DoF = 370/384, providing a reasonably good improvement over the
\diskpbb\ case ($\Delta\chi^{2} = 13$ for one additional free parameter). The parameter
constraints for this model are also presented in Table \ref{tab_5907_phaseres}, and we
show the relative contribution of the model components for the peaks and the troughs
of the pulse cycle in Figure \ref{fig_5907_phaseres}. The strong degeneracy seen with
the \diskpbb\ model is not present here, as neither the \diskbb\ or BB models can mimic
the \cutoffpl\ component, and the temperatures of the two thermal components ($kT
\sim 0.4$ and $\sim$1.3\,keV) are broadly similar to those seen with the same model in
\p13\ ($kT \sim 0.3-0.5$ and $\sim$1.1$-$1.5\,keV). Here, we clearly see that the
\cutoffpl\ component associated with the accretion column dominates the highest
observed energies.

\section{The Broadband ULX Sample: Hard Excesses}
\label{sec_sample_hardex}

Having established that the hard excesses in both \p13\ and \ngc\ are associated with
emission from the accretion columns (see also \citealt{Walton18p13}), in this section
we assess the presence of similar features in the rest of the ULX population. For this 
analysis, we focus on the sample with high S/N broadband observations highlighted in
\cite{Walton18p13}. These are sources for which \nustar\ has performed simultaneous
observations with \xmm\ and/or\ \suzaku\ (\citealt{SUZAKU}) resulting in a robust
detection up to at least 20\,keV, and which are not significantly confused with any other
X-ray sources. In addition to the neutron star ULXs \ngc\ and \p13, the sample consists
of: Circinus ULX5 (\citealt{Walton13culx}), Holmberg II X-1 (\citealt{Walton15hoII}),
Holmberg IX X-1 (\citealt{Walton14hoIX, Walton17hoIX, Luangtip16}), IC\,342 X-1 and
X-2 (\citealt{Rana15}), NGC\,1313 X-1 (\citealt{Bachetti13, Miller14, Walton16ufo}) and
NGC\,5204 X-1 (\citealt{Mukherjee15}). Details of all the observations considered in
this work are given in Appendix \ref{app_obslog}. We do not include \m82\ because,
although we can isolate the pulsed emission, we cannot extract clean spectra of this
source owing to its proximity to M82 X-1 (also absent from our sample). M82 X-1 is
typically much brighter than X-2, reaching X-ray luminosities of up to
$\sim$10$^{41}$\,\ergps\ (\eg\ \citealt{Kaaret09m82, Brightman16m82b}).

Of the sources for which the nature of the accretor still remains unknown, hard
excesses have already been reported in the literature for Circinus ULX5, Holmberg II
X-1, Holmberg IX X-1 and NGC 5204 X-1 when the lower-energy data have been fit
with thermal accretion disk models. Here, we assess whether similar features would
also be required in the remaining sources, IC\,342 X-1 and X-2 and NGC\,1313 X-1,
when fit with similar models. Since pulsations have not been detected for any
of these sources, we fit their time-averaged spectra with an accretion disk model
combining \diskbb+\diskpbb, allowing for an outer thin disk (exterior to \rsp;
\diskbb) and an inner thick disk (\diskpbb; note that this is a slightly
different model combination than the previous section, but has been frequently used in
previous works). Neutral absorption is included as before (both Galactic and intrinsic;
see Table \ref{tab_hardex_sample} for the Galactic absorption columns). We then test
whether these models require an additional high-energy continuum component by
determining whether adding a phenomenological powerlaw tail to the hotter \diskpbb\
component (using the \simpl\ model; \citealt{SIMPL}) provides a
significant improvement to the fit.

\begin{figure}
\begin{center}
\hspace*{-0.6cm}
\rotatebox{0}{
{\includegraphics[width=235pt]{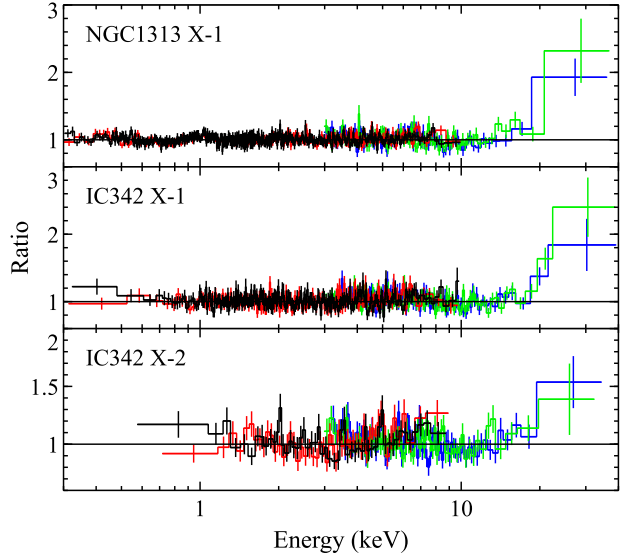}}
}
\end{center}
\caption{
Data/model ratios to the DISKBB+DISKPBB continuum model for the ULXs NGC\,1313
X-1 (top), IC\,342 X-1 (middle) and IC\,342 X-2 (bottom). The data from \epicpn,
\epicmos, FPMA and FPMB are shown in black, red, green and blue, respectively. In
all three cases, excess emission is seen at the highest energies probed by \nustar. In
the case of NGC\,1313 X-1, the residuals at 1\,keV are caused by the ultrafast outflow
known in this source (\citealt{Middleton15, Pinto16nat}), and do not influence the
presence of the hard excess.}
\label{fig_hardex}
\vspace{0.3cm}
\end{figure}

Figure \ref{fig_hardex} shows the residuals for the \diskbb+\diskpbb\ models for the
three sources. Excesses are seen at the highest energies in all three cases, similar to
those seen in the rest of the population, and we find that the addition of the \simpl\
component provides a significant improvement to the fit with the \diskbb+\diskpbb\
continuum ($\Delta\chi^{2} > 30$ for two additional DoF). For NGC\,1313 X-1, we
stress that this is the case regardless of whether the atomic emission and absorption
features associated with the ultrafast outflow reported in \cite{Pinto16nat} and
\cite{Walton16ufo} (which produce the residuals seen at $\sim$1\,keV) are included in
the model. We summarise the presence of these hard excesses in the broadband
ULX population in Table \ref{tab_hardex_sample}. All the ULXs with broadband
observations to date require an additional high-energy component when fit with
dual-thermal disk models that may represent a super-Eddington accretion flow.

\begin{table*}
  \caption{Summary of the hard excess detections when in the ULX population with broadband coverage to date when the lower-energy data are fit with thermal disk
  models}
  \vspace{-0.3cm}
\begin{center}
\begin{tabular}{c c c c c c c c}
\hline
\hline
\\[-0.1cm]
Source & Distance & Reference & $N_{\rm{H; Gal}}$ (K05) & Hard Excess & Reference & Accretor \\
\\[-0.2cm]
& [Mpc] & & [$10^{20}$\,\pcmsq] & Detected? & \\
\\[-0.2cm]
\hline
\hline
\\[-0.1cm]
Circinus ULX5 & 4.2 & F77 & 55.8 & Yes & W13 & Unknown \\
\\[-0.2cm]
Holmberg II X-1 & 3.4 & K02, T16 & 3.66 & Yes & W15 & Unknown \\
\\[-0.2cm]
Holmberg IX X-1 & 3.6 & P02, T16 & 5.54 & Yes & W14, L16, W17b & Unknown \\
\\[-0.2cm]
IC\,342 X-1 & 3.4 & S02, T16 & 29.9 & Yes & This work & Unknown \\
\\[-0.2cm]
IC\,342 X-2 & 3.4 & S02, T16 & 29.9 & Yes & This work & Unknown \\
\\[-0.2cm]
NGC\,1313 X-1 & 4.2 & M02, T16 & 4.13 & Yes & This work & Unknown \\
\\[-0.2cm]
NGC\,5204 X-1 & 4.9 & T16 & 1.75 & Yes & M15 & Unknown \\
\\[-0.2cm]
NGC\,5907 ULX1 & 17.1 & T16 & 1.21 & Yes & F17, this work & Neutron star \\
\\[-0.2cm]
NGC\,7793 P13 & 3.5 & P10, T16 & 1.20 & Yes & W17a & Neutron star \\
\\[-0.2cm]
\hline
\hline
\\[-0.35cm]
\end{tabular}
\label{tab_hardex_sample}
\end{center}
\flushleft
References: F77 -- \cite{Freeman77}, F17 -- \cite{Fuerst17ngc5907}, K02 --
\cite{Karachentsev02}, K05 -- \cite{NH}, L16 -- \cite{Luangtip16}, M02 --
\cite{Mendez02}, M15 -- \cite{Mukherjee15}, P02 -- \cite{Paturel02}, P10 --
\cite{Pietrzynski10}, S02 -- \cite{Saha02}, T16 -- \cite{Tully16}, W13-17b --
\cite{Walton13culx, Walton14hoIX, Walton15hoII, Walton18p13, Walton17hoIX}
\vspace{0.3cm}
\end{table*}

\section{The Broadband ULX Sample: ULX Pulsar Fits}
\label{sec_sample_ULXp}

We finish our analysis by testing whether the average spectra of the broadband ULX
population can be well fit with a model similar to that applied to both \p13\ and \ngc,
continuing to focus on the broadband ULX sample discussed in Section
\ref{sec_sample_hardex} for illustration. In particular, having demonstrated that even
complex accretion disk models require an additional continuum component at the
highest energies probed in all ULXs with broadband observations to date (Table
\ref{tab_hardex_sample}), we wish to test  whether these excesses can all be
explained with emission from an accretion column similar to those seen in the three
known ULX pulsars. 

This is similar in concept to the recent works by \cite{Pintore17} and \cite{Koliopanos17}.
The former fit a model commonly applied to sub-Eddington pulsars to a sample of
time-averaged \xmm\ and \nustar\ ULX spectra, undertaking a broadband analysis where
possible, and the latter fit a model motivated by the model for ULX pulsars proposed by
\citet[][see below]{Mushtukov17} to another sample of time-averaged ULX spectra, but
do not undertake a full broadband analysis, instead treating the soft and hard X-ray data
from \xmm\ and \nustar\ separately. We stress that our analysis is strictly broadband (the
soft and hard X-ray data are analysed simultaneously), and the model used here is
directly motivated by our observational analysis of the two ULX pulsars for which clean
broadband spectroscopy is possible, and which have super-Eddington luminosities.

The model applied here consists of two thermal blackbody components for the accretion
flow beyond \rmag, and a \cutoffpl\ component for the accretion column, as required to
fit the broadband data from \p13\ (\citealt{Walton18p13}) and also preferred by the data
for \ngc\ (Section \ref{sec_5907_phaseres}). As before, we allow for both Galactic
and intrinsic neutral absorption. For the two known ULX pulsars in the sample, \p13\ and
\ngc, the shape parameters for the \cutoffpl\ component are fixed to the best-fit values
obtained for their pulsed emission ($\Gamma = 0.17$, $E_{\rm{cut}} = 4.7$\,keV and
$\Gamma = 0.7$, $E_{\rm{cut}} = 5.5$\,keV, respectively; Figure \ref{fig_pulse_gamEc}).
However, pulsations have not been detected for the majority of the sample, so we
cannot isolate the emission from any accretion column present in these systems. In
these cases, we set the shape parameters to the average values seen from the pulsed
emission from the three ULX pulsars currently known: $\Gamma = 0.5$ and $E_{\rm{cut}}
= 8.1$\,keV (also shown in Figure \ref{fig_pulse_gamEc}).

For the thermal components, we construct a simple decision tree to determine which
model components to fit in a systematic manner. Based on our analysis of \p13,
we assume that \rmag\ $<$ \rsp, such that the thick inner disk can form before being
truncated by the magnetic field. Initially, we therefore make use of the \diskbb+\diskpbb\
combination for the reasons discussed above (to remove any degeneracy
between these components and prevent them from swapping temperatures, we set an
upper limit for the \diskbb\ temperature of 1\,keV in our fits). However, as
discussed in \cite{Walton18p13}, if \rsp\ and \rmag\ are similar, the thick inner disk may
only extend over a small range of radii before being truncated by the magnetic field of
the neutron star, and subsequently only emit over a relatively narrow range of
temperatures. If this is the case then we should see a steeper radial temperature index
(\ie $p > 0.75$), as the \diskpbb\ model implicitly assumes that the emission
from this component extends out to large radii (and therefore low temperatures), so
increasing $p$ is the only way the model can force the hotter component to be
dominated by a small range of temperatures. In this case, the hotter component may be
better described by a single blackbody (as appears to be the case for both \p13\ and
\ngc) than a disk component with a broad range of temperatures. Therefore, if we see
that $p$ runs up against its upper limit in fits with the \diskpbb\ component (we
restrict the radial temperature index to the range $0.5 \leq p \leq 2.0$), we replace it
with a BB component and present these fits instead.

\begin{table*}
  \caption{Best fit parameters obtained for the fits to the broadband ULX sample with ULX pulsar models}
\vspace{-0.3cm}
\begin{center}
\hspace*{-0.15cm}
\begin{tabular}{c c c c c c c c c c c}
\hline
\hline
\\[-0.1cm]
Source & Thermal & $N_{\rm H; int}$ & $kT_{1}$ & Norm$_{1}$ & $kT_{2}$ & $p$ & Norm$_{2}$\tmark[b] & CPL & $\chi^{2}$/DoF \\
\\[-0.2cm]
& Continuum\tmark[a] & [$10^{20}$\,cm$^{-2}$] & [\kev] & & [\kev] & & & Flux\tmark[c] \\
\\[-0.15cm]
\hline
\hline
\\[-0.1cm]
N7793 P13 & DBB+BB & $7.1^{+0.8}_{-0.7}$ & $0.48 \pm 0.03$ & $0.80^{+0.19}_{-0.15}$ & $1.53 \pm 0.04$ & -- & $27.3^{+2.4}_{-2.2}$ & $22.3^{+1.4}_{-1.6}$ & 1132/1159 \\
\\[-0.1cm]
Circ. ULX5 & DPBB & $21 \pm 3$ & -- & -- & $1.87 \pm 0.07$ & $0.71 \pm 0.02$ & $24^{+6}_{-4}$ & $6.7^{+1.2}_{-1.5}$ & 1216/1147 \\
\\[-0.1cm]
Ho II X-1 & DBB+DPBB & $4.7^{+1.2}_{-1.0}$ & $0.24^{+0.02}_{-0.01}$ & $32^{+11}_{-7}$ & $2.07^{+0.14}_{-0.12}$ & $0.56 \pm 0.02$ & $3.6^{+1.5}_{-1.0}$ & $4.7^{+0.6}_{-0.7}$ & 1954/1965 \\
\\[-0.1cm]
Ho IX X-1 (L)\tmark[d] & DBB+DPBB & $16 \pm 1$ & $0.31^{+0.03}_{-0.04}$ & $9^{+6}_{-3}$ & $2.7^{+0.5}_{-0.3}$ & $0.63^{+0.08}_{-0.04}$ & $2.4^{+3.2}_{-1.3}$ & $20.3^{+2.9}_{-5.4}$ & 6268/5901 \\
\\[-0.1cm]
Ho IX X-1 (M)\tmark[d] & DBB+DPBB & -- & $0.27 \pm 0.02$ & $12^{+7}_{-4}$ & $3.0 \pm 0.2$ & $0.54 \pm 0.01$ & $1.4^{+0.4}_{-0.3}$ & $20.0^{+2.0}_{-1.3}$ & -- \\
\\[-0.1cm]
Ho IX X-1 (H)\tmark[d] & DBB+DPBB & -- & $0.20^{+0.04}_{-0.03}$ & $36^{+51}_{-21}$ & $1.92 \pm 0.04$ & $0.63 \pm 0.01$ & $34 \pm 4$ & $20.2^{+1.0}_{-1.2}$ & -- \\
\\[-0.1cm]
IC\,342 X-1 & DBB+DPBB & $60^{+8}_{-5}$ & $0.36^{+0.04}_{-0.06}$ & $3.6^{+4.8}_{-1.3}$ & $2.1^{+0.3}_{-0.2}$ & $0.63^{+0.08}_{-0.07}$ & $3.2^{+2.8}_{-1.9}$ & $8.3^{+0.5}_{-1.1}$ & 1351/1308 \\
\\[-0.1cm]
IC\,342 X-2 & BB & $90^{+7}_{-6}$ & -- & -- & $1.28 \pm 0.05$ & -- & $9.8 \pm 0.6$ & $10.8^{+0.5}_{-0.4}$ & 672/659 \\
\\[-0.1cm]
N1313 X-1 & DBB+DPBB & $22 \pm 1$ & $0.29 \pm 0.01$ & $11.4^{+2.1}_{-0.9}$ & $2.7^{+0.2}_{-0.3}$ & $0.58^{+0.02}_{-0.01}$ & $1.1^{+0.6}_{-0.4}$ & $4.6^{+0.8}_{-1.2}$ & 1840/1691 \\
\\[-0.1cm]
N5204 X-1 & DBB+DPBB & $2.8^{+1.5}_{-1.4}$ & $0.27^{+0.02}_{-0.03}$ & $7.1^{+3.4}_{-2.1}$ & $1.8 \pm 0.2$ & $0.64^{+0.11}_{-0.06}$ & $2.8^{+3.7}_{-1.4}$ & $1.8^{+0.4}_{-0.5}$ & 548/520 \\
\\[-0.1cm]
N5907 ULX1 (L)\tmark[d] & DBB+BB & $66^{+8}_{-2}$ & $0.46^{+0.07}_{-0.09}$ & $0.39^{+0.70}_{-0.17}$ & $1.19^{+0.14}_{-0.19}$ & -- & $3.2 \pm 0.8$ & $2.6^{+0.7}_{-0.8}$ & 598/646 \\
\\[-0.1cm]
N5907 ULX1 (H)\tmark[d] & DBB+BB & -- & $0.50^{+0.09}_{-0.11}$ & $0.30^{+0.76}_{-0.15}$ & $1.31^{+0.14}_{-0.19}$ & -- & $4.2^{+1.5}_{-1.4}$  & $13.5 \pm 1.2$ & -- \\
\\[-0.2cm]
\hline
\hline
\\[-0.3cm]
\end{tabular}
\label{tab_param_sample}
\end{center}
\flushleft
$^a$ DBB = DISKBB, DPBB = DISKPBB, BB = blackbody. Where two thermal
components are used, the DISKBB is always the cooler of the two ($kT_{1}$,
Norm$_{1}$), with its temperature limited to be $<$1\,keV to reduce degeneracy
between the components. \\
$^b$ For the DISKPBB model the normalisation is given in units of $10^{-3}$, and
for the BB model it is given in units of $10^{-6}$. \\
$^c$ CPL = CUTOFFPL. The shape parameters for this component are fixed (see
Section \ref{sec_sample_ULXp}), and the 2-10\,keV fluxes are given in units of
$10^{-13}$\,\ergpcmsqps. \\
$^d$ For sources where multiple observations are considered, we assume a common
$N_{\rm H; int}$ across all epochs; the \chisq\ quoted is for the combined fit.
\end{table*}

\begin{figure*}
\begin{center}
\rotatebox{0}{
{\includegraphics[width=490pt]{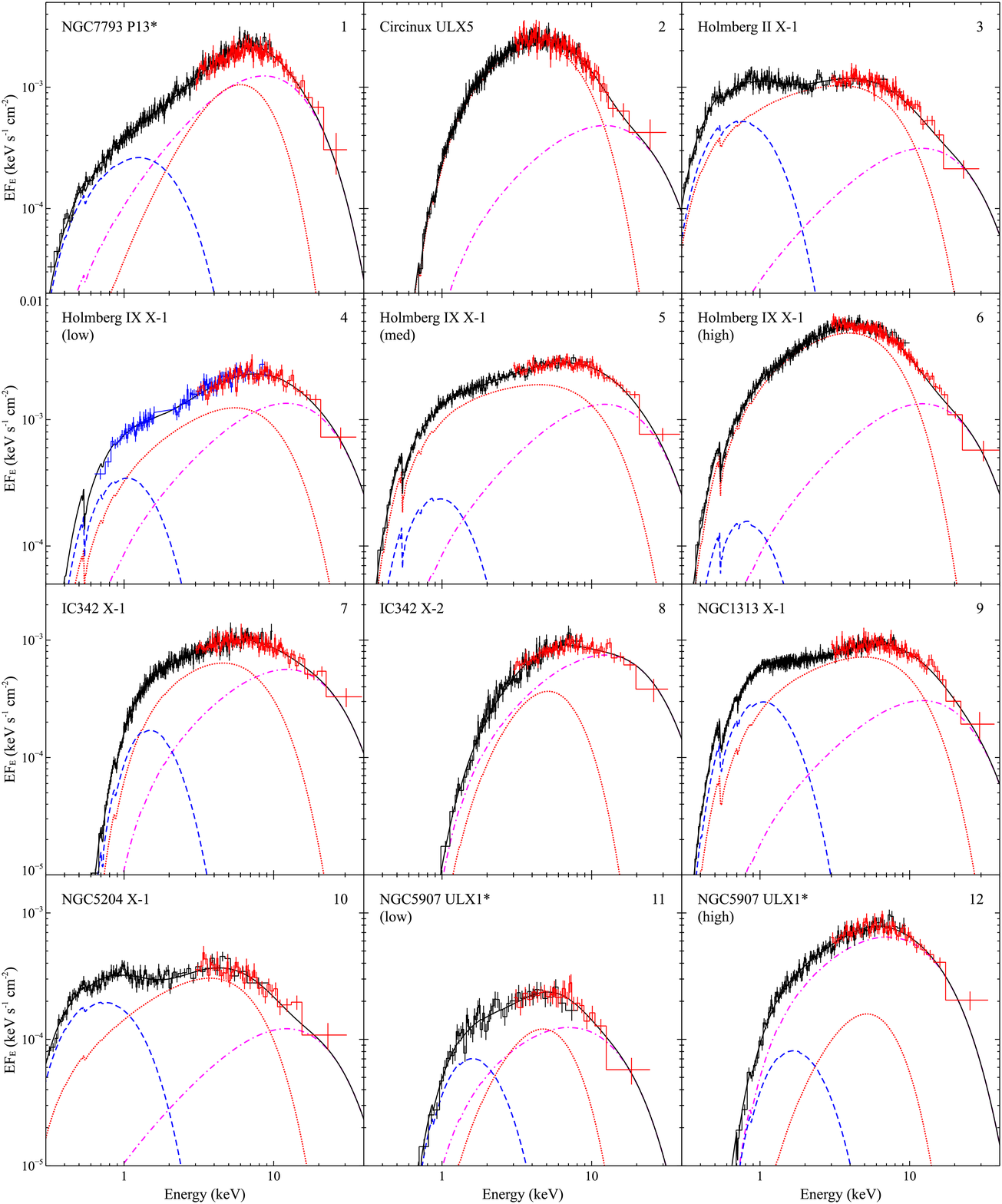}}
}
\end{center}
\caption{Broadband spectra of the ULX population with simultaneous coverage in soft
(\xmm, \suzaku) and hard (\nustar) X-rays, and the relative contributions of the various
model components included in our fits based on the known ULX pulsars (indicated with
an asterisk). We show the same observations as Figure 11 from \citet[as before, \xmm\
\epicpn\ data is in black, \suzaku\ front-illuminated XIS is in blue, and \nustar\ data is in
red]{Walton18p13}, and keep the same ordering of the panels for ease of comparison,
but here the data have been unfolded through the best-fit model for each source (solid
black line). These models generally include two thermal components likely from the
accretion flow beyond \rmag, with the cooler shown in dashed blue (modelled with 
DISKBB; note that on occasion this component is not required) and the hotter shown in
dotted red (modelled with either DISKPBB or a single blackbody), and a high-energy
component representing emission from a ULX pulsar-like accretion column, which is
shown in dash-dot magenta (modelled with a CUTOFFPL component); see Section
\ref{sec_sample_ULXp} for individual source details. Where multiple observations of
the same source are shown, the relative fluxes of these datasets are indicated in
parentheses.}
\label{fig_sample}
\end{figure*}

\begin{figure*}
\begin{center}
\vspace*{0.5cm}
\rotatebox{0}{
{\includegraphics[width=490pt]{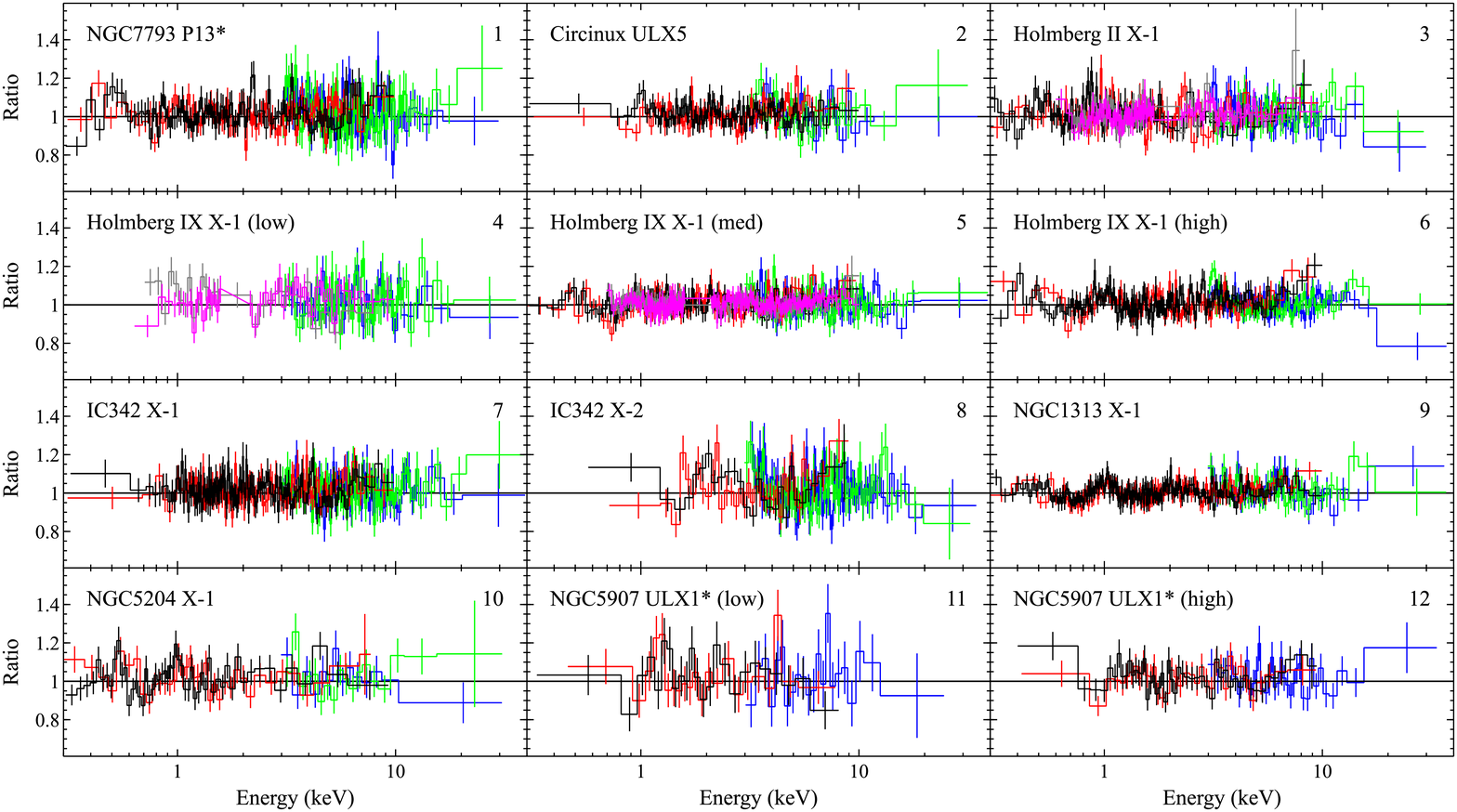}}
}
\end{center}
\caption{
Data/model ratios for our ULX pulsar fits to the broadband ULX sample (see Section
\ref{sec_sample_ULXp}); the panel ordering and source/observation nomenclature
are the same as Figure \ref{fig_sample}. Here, we show the full dataset fit in each case,
with data from \epicpn\ and \epicmos\ (both \xmm), the front- and back-illuminated XIS
(both \suzaku) and FPMA and FPMB (both \nustar) shown in black, red, magenta, grey,
green and blue, respectively (in the case of \ngc, the combined FPMA+FPMB data are
shown in blue). The data have been further rebinned for visual purposes.}
\label{fig_sample_ratio}
\end{figure*}

Lastly, to keep the models as simple as possible and minimize parameter degeneracies,
we test to see whether both the thermal components are required by the data. Should
the fit be similarly good after removing either of the thermal components included in the
model, then we present these fits instead, retaining the component that provides the
better fit of the two. However, where multiple epochs are considered for the same
source (Holmberg IX X-1, \ngc) we also make sure to use the same model for all epochs
for consistency, such that if one observation prefers two thermal components, while
another does not, we still use the same two thermal components for all observations.
For these sources, we also fit all of the observations considered simultaneously,
allowing us to adopt a common neutral absorption column (\eg \citealt{Miller13ulx,
Walton17hoIX}). We stress, however, that fully consistent results are obtained allowing
the column to vary between epochs.

The results are presented in Table \ref{tab_param_sample}, and we show the relative
contributions of the model components for each of the spectra analysed here in
Figure \ref{fig_sample}. We also show the data/model ratios for the fits in Figure
\ref{fig_sample_ratio}. For \p13, since the model applied here is identical to that used
in our recent work focusing on this individual source, we take the results from
\cite{Walton18p13}. Good broadband fits to the data are obtained in all cases, and
critically we find that emission from a ULX pulsar-like accretion column can reproduce
the hard X-ray \nustar\ data for all the sources for which the nature of the accretor
currently remains unknown.

For most of the datasets considered, we fit the thermal continuum with the
\diskbb+\diskpbb\ combination, and the radial temperature indices for the hotter
\diskpbb\ component are flatter than expected for a thin disk. The exceptions are the
two ULX pulsars (\p13, \ngc), Circinus ULX5 and IC\,342 X-2. For both the ULX
pulsars, we find that when using the \diskpbb\ model only a lower limit can be
obtained on $p$, so we revert to a single blackbody for the hotter component. The
only other source for which this is the case is IC\,342 X-2, where we similarly switch
from \diskpbb\ to a single BB. Here, we also find that the cooler \diskbb\ component
is not required by the data, but we note that this is the most absorbed of the sources
considered ($N_{\rm{H; int}} \sim 10^{22}$\,\pcmsq), so the fact that this component
is not required could merely be related to the high levels of absorption. For Circinus
ULX5, we also find that the \diskbb\ component is not required, but again this source
is rather absorbed as the Galactic column towards Circinus is quite high ($N_{\rm{H;
Gal}} \sim 5.6 \times 10^{21}$\,\pcmsq, giving a total column of $\sim$8 $\times
10^{21}$\,\pcmsq). However, in this case, we still find that the \diskpbb\ component
prefers a fairly flat radial temperature index, similar to the bulk of the other datasets.

\section{Discussion}
\label{sec_dis}

The relative contributions of black hole and neutron star accretors to the ULX
population is a matter of some debate. All three of the neutron star ULXs currently
known have been discovered through the detection of coherent X-ray pulsations.
Although simple searches have now been made for a broad section of the ULX
population (\eg\ \citealt{Doroshenko15}, Appendix \ref{app_pulse}), pulsations have
not been seen from any other ULX to date. However, these signals can be difficult to
find. Being extragalactic, ULXs are typically faint, resulting in low count rates. In
addition, motions of the neutron star within its binary orbit (the parameters of which
are not generally known a priori) can shift the frequency the pulsations should be
observed at across the duration of individual observations. Both of these issues can
hinder the detection of pulsations, even if they are intrinsically present. Furthermore,
in 2/3 of the ULX pulsars known, where we know the pulsations are detectable and
constraints on the binary orbit have subsequently been obtained, the pulsations are
transient. In \m82, for example, the duty cycle of the pulsations is $<$50\%. Additional
means of identifying neutron star ULXs beyond the detection of pulsations may
therefore be key to addressing their contribution to the overall ULX population.

\cite{Walton18p13} discuss the qualitative similarity of the broadband spectra of
all the ULXs observed with clean, high S/N broadband spectra, \ie combining \nustar\
with either \xmm\ and/or \suzaku. This sample consists of known ULX pulsars
(\p13\ and \ngc), as well as ULX with unknown accretors (Circinus ULX5, Holmberg\,II
X-1, Holmberg\,IX X-1, IC\,342 X-1 and X-2, NGC\,1313 X-1 and NGC\,5204 X-1; see
Section \ref{sec_sample_hardex}). Here, we have demonstrated that \textit{all} these
sources require an additional high-energy continuum component that dominates above
$\sim$10\,keV when the lower energy data is modelled with thermal accretion disk
models. We note that, based on their analysis of \nustar\ data for a number of the
sources considered here, \cite{Koliopanos17} suggest that the need for this additional
component is an artefact of mis-modelling the hotter ($\sim$2--3\,keV) thermal
emission. Fitting the 4--30\,keV data with a simple \diskbb\ model they find that an
additional high-energy component would be required, but after switching to the more
complex \diskpbb\ model they find that this is no longer the case. However, this is
clearly a consequence of the limited bandpass considered by these authors; the
\diskpbb\ model has routinely been used to fit this thermal emission, both in this work
and in the existing literature (\eg\ \citealt{Mukherjee15, Walton15hoII, Walton17hoIX}),
and when fitting the full $\sim$0.3--30\,keV broadband data an additional high-energy
component is still required.

In both NGC\,7793 P13 and NGC\,5907 ULX1, based on phase-resolved analyses
we find that the additional high-energy emission is pulsed and therefore associated
with the magnetically collimated accretion columns that must be present in these
sources. Although we cannot separate the broadband X-ray emission from M82 X-1
and X-2, and as such these sources are not in our sample, we do note that the
pulsed emission from \m82\ also has a very similar spectral form as \p13\ and \ngc\
(\citealt{Brightman16m82a}), with significant emission in high-energy X-rays (and we
also note that there is some evidence for a hard excess in M82 X-1 as well;
\citealt{Brightman16m82b}).

If the presence of a hard excess is an indication that the central accretion operates in
this manner, then the fact that the entire sample shows this feature is consistent with
them all hosting neutron star accretors. Indeed, fitting the broadband data with a
model inspired by that required to fit the known ULX pulsars, in which the highest
energy emission probed by \nustar\ is dominated by an accretion column similar to
those seen in the ULX pulsar systems, provides a good fit to the available high-energy
data in all cases.

\subsection{Accretion Geometry and the Lack of Pulsations}

Our fits suggest why, even if they are all neutron stars, pulsations have not been seen
in any cases other than \p13\ and \ngc\ among the sample considered here (\eg\
\citealt{Doroshenko15}, Appendix \ref{app_pulse}). The final model applied to the
broadband data here (typically) consists of two thermal components, which we assume
arise in the accretion flow beyond \rmag\ (with the cooler component arising in the
regions of the disk beyond \rsp, and the hotter component the regions between \rsp\
and \rmag), and a high-energy continuum (modelled with a \cutoffpl\ component) from
the accretion column. Although the accretion column is always required in our fits to
account for the hard excesses in the data, from Figure \ref{fig_sample} it is clear that
among the rest of the sample this generally makes a smaller relative contribution to
the total emission than for the two observations of the known ULX pulsars from which
pulsations have been detected, sometimes substantially so. To quantitatively illustrate
this, we summarise the total 0.3--40.0\,keV observed fluxes ($F_{\rm{tot}}$) and the
fluxes inferred for the \cutoffpl\ component alone ($F_{\rm{col}}$) in Table
\ref{tab_fluxcomp}, and highlight the observations where pulsations have been
detected. We stress that these fits are only supposed to be illustrative, since we do not
actually know the precise spectral form of any accretion columns present in the
remaining sample, and substantial deviations from our assumed shape could lead to
larger errors than the simple statistical errors computed given this shape (see below).
Nevertheless, the fluxes confirm the visual conclusion from Figure \ref{fig_sample}; for
the observations in which pulsations have been detected our analysis suggests that
that the column has provided $\sim$60\% or more of the total observed flux, while for
sources in which pulsations have not currently been detected we generally infer much
lower values than this. The pulse fractions for the majority of the rest of the sample
would therefore be diluted in comparison to both \p13\ and \ngc, and therefore any
pulsations would be harder to detect even before any S/N considerations.

\begin{table}
  \caption{A comparison of the the total observed flux and the flux from the
  observed/potential accretion columns in the 0.3--40\,keV band based on the ULX
  pulsar fits for the datasets considered in this work.}
\vspace{-0.4cm}
\begin{center}
\begin{tabular}{c c c c c}
\hline
\hline
\\[-0.1cm]
Dataset & $F_{\rm{tot}}$ & $F_{\rm{col}}$ & $F_{\rm{col}}/F_{\rm{tot}}$ \\
\\[-0.2cm]
& \multicolumn{2}{c}{[$10^{-12}$\,\ergpcmsqps]} \\
\\[-0.2cm]
\hline
\hline
\\[-0.1cm]
Circ ULX5 & $73.5 \pm 1.3$ & $14.8^{+2.6}_{-3.6}$ & $0.20^{+0.03}_{-0.04}$ \\
\\[-0.2cm]
Ho II X-1 & $60.4 \pm 0.9$ & $10.4^{+1.4}_{-1.6}$ & $0.17^{+0.02}_{-0.03}$ \\
\\[-0.2cm]
Ho IX X-1 (L) & $91.6^{+2.2}_{-2.3}$ & $44.3^{+6.4}_{-11.7}$ & $0.48^{+7}_{-13}$ \\
\\[-0.2cm]
Ho IX X-1 (M) & $123.0 \pm 1.4$ & $43.4^{+4.3}_{-5.8}$ & $0.35^{+0.04}_{-0.05}$ \\
\\[-0.2cm]
Ho IX X-1 (H) & $199.5 \pm 1.8$ & $43.9^{+2.4}_{-2.5}$ & $0.22 \pm 0.01$ \\
\\[-0.2cm]
IC\,342 X-1 & $37.9 \pm 0.6$ & $17.8^{+1.1}_{-2.2}$ & $0.47^{+0.03}_{-0.06}$ \\
\\[-0.2cm]
IC\,342 X-2 & $30.7 \pm 0.8$ & $23.4 \pm 1.0$ & $0.76 \pm 0.04$ \\
\\[-0.2cm]
N1313 X-1 & $40.2 \pm 0.5$ & $10.0^{+1.8}_{-2.6}$ & $0.25^{+0.04}_{-0.07}$ \\
\\[-0.2cm]
N5204 X-1 & $18.5 \pm 0.5$ & $4.1^{+0.7}_{-1.2}$ & $0.22^{+0.04}_{-0.06}$ \\
\\[-0.2cm]
N5907 ULX1 (L) & $7.7 \pm 0.4$ & $4.1^{+0.9}_{-1.3}$ & $0.53^{+0.13}_{-0.17}$ \\
\\[-0.2cm]
N5907 ULX1 (H)* & $25.7^{+0.7}_{-0.6}$ & $21.2^{+1.7}_{-2.0}$ & $0.82^{+0.07}_{-0.08}$ \\
\\[-0.2cm]
N7793 P13* & $67.0 \pm 1.1$ & $39.7^{+2.6}_{-2.9}$ & $0.59 \pm 0.04$ \\
\\[-0.2cm]
\hline
\hline
\\[-0.45cm]
\end{tabular}
\label{tab_fluxcomp}
\end{center}
* Datasets for which pulsations have been detected (note that pulsations were only
detected in the higher flux observation of NGC\,5907 ULX1)
\vspace{0.3cm}
\end{table}

\begin{figure*}
\begin{center}
\vspace*{0.5cm}
\rotatebox{0}{
{\includegraphics[width=425pt]{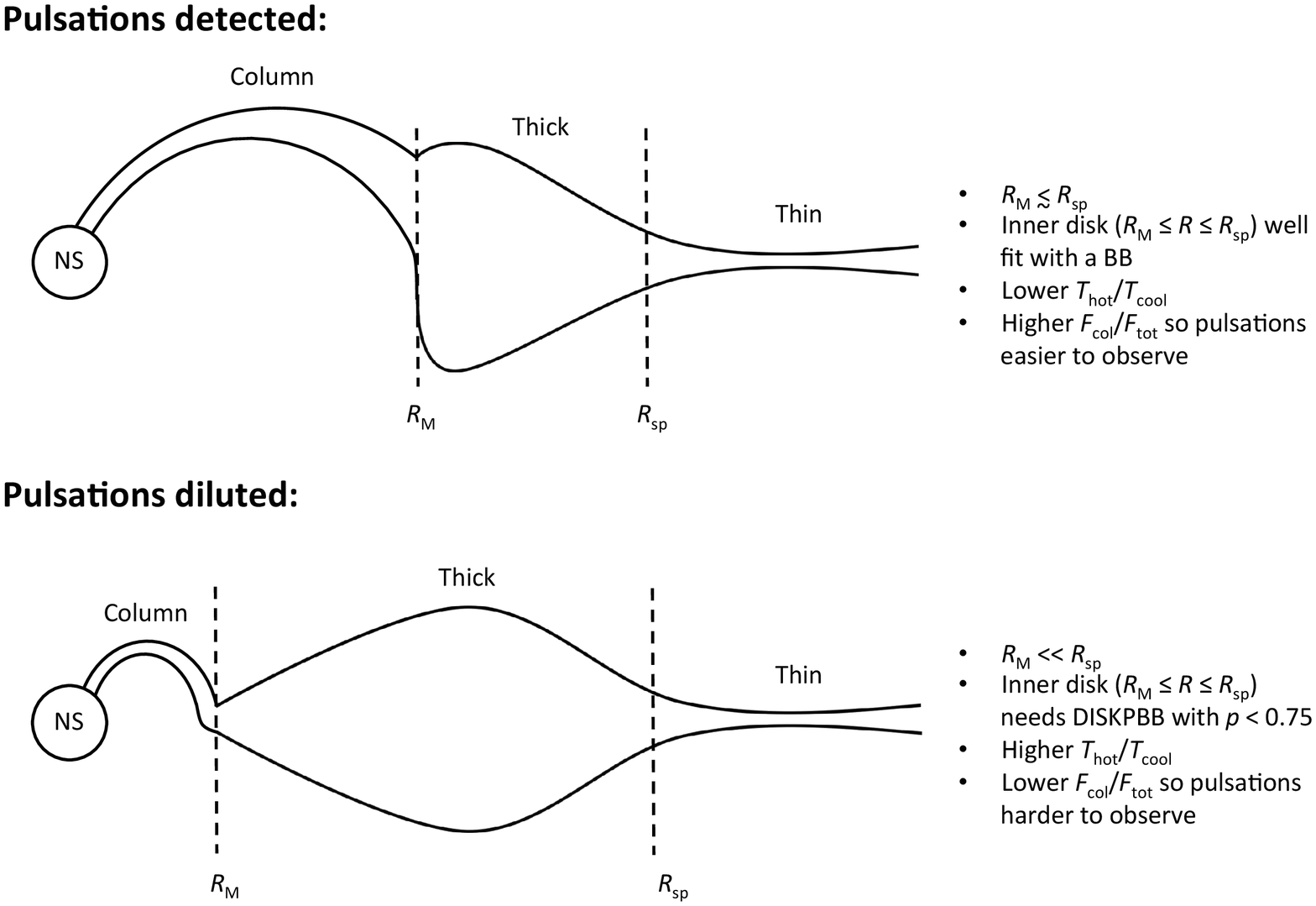}}
}
\end{center}
\vspace{-0.2cm}
\caption{
Schematic diagram illustrating the potential difference between the pulsar ULXs (top)
and the ULXs for which pulsations have not been detected (bottom), assuming neutron
star accretors (not to scale). In the former case, the magnetic field of the neutron star
truncates the disk relatively close to the spherization radius, resulting in the thick inner
disk spanning a narrow range of radii and a stronger relative contribution from the
emission from the accretion column, making pulsations easier to detect. In the latter
case, the disk is truncated at radii much smaller than the spherization radius, allowing
the thick inner disk to span a broad range of radii and resulting in a weaker relative
contribution from the accretion column, in turn making pulsations harder to detect. In
this case, the structure of the disk should be very similar to that calculated for the black
hole case (\eg\ \citealt{Poutanen07, Dotan11}).}
\label{fig_schematic}
\vspace{0.3cm}
\end{figure*}

The only exception is IC\,342 X-2, where we would infer a relative contribution from
the column to the total flux similar to the known pulsar ULXs. Is it also interesting to
note that this is the only other source for which we find the \diskpbb\ component
prefers a radial temperature index steeper than a standard thin disk, so that it can be
replaced with a single blackbody (see Section \ref{sec_sample_ULXp}). In
\cite{Walton18p13} we suggested that this could be an indication that the magnetic
field of the neutron star truncates the disk fairly close to the point that it becomes
locally Eddington (\ie \rmag\ $\sim$ \rsp), such that the thick inner disk only extends
over a small range of radii and subsequently emits over a narrow range of
temperatures. This was also supported by the rough similarity (within a factor of
$\sim$5) of the characteristic radii estimated for both the \diskbb\ and BB components
for \p13\, and we speculated that this might therefore be an indicator of a neutron
star accretor since such truncation would not be possible for black hole systems.
\cite{King17ulx} suggest that \rmag\ $\sim$ \rsp\ may be a necessary requirement for
the detection of pulsations, since the steady (\ie non-pulsed) emission from the
accretion flow beyond \rmag\ dilutes the pulsations less, consistent with our \p13\
results. We note again that IC\,342 X-2 is the most absorbed of the sources
considered, with a rather substantial column density ($\sim$10$^{22}$ \pcmsq), and
we have been forced to assume a spectral form for the accretion column. In
combination, these could potentially lead to large uncertainties in our continuum fits
(see below). Nevertheless, although nothing has been detected to date, IC\,342 X-2
might therefore be an interesting source for future pulsation searches.

However, we stress that for the other sources, where we obtain flatter radial
temperature indices ($p < 0.75$), this would not exclude neutron star accretors. We
expect that this would suggest that \rmag\ $\ll$ \rsp\ in these cases, such that the thick
inner disk extends over a sufficiently large range of radii/temperatures that the model
returns values in line with the broad expectations for such an accretion flow. This
would potentially explain why the accretion column appears to be weaker in a relative
sense in these cases, as the flow beyond \rmag\ dominates more of the observed
emission, and is qualitatively consistent with the above suggestion from \cite{King17ulx}
given the lack of pulsations detected from these systems. In order for \rmag\ $\ll$ \rsp,
either the accretion rate ($\dot{M}$) must be substantially higher, or the magnetic field
substantially weaker in these systems in comparison to those where \rmag\ $\sim$ \rsp,
as \rmag\ $\propto \dot{M}^{-2/7} B^{4/7}$ (\citealt{Lamb73, Cui97})\footnote{We note
the caveat that these models assume the disk outside \rmag\ is thin, which may not be
the case in these systems. However, we expect that the qualitative dependencies
between \rmag\ and $\dot{M}$ and $B$ (\ie higher $B$ resulting in a larger \rmag) are
likely robust to this issue.} and \rsp\ $\propto \dot{M}$ (\citealt{Shakura73}). A rough
schematic diagram of the potential difference between the known pulsar ULXs and
sources for which pulsations have not been detected is shown in Figure
\ref{fig_schematic}.

\citealt{Mushtukov17} and \cite{Koliopanos17} discuss an alternative scenario. In their
picture, the hotter thermal component arises in the accretion `curtain' formed just
interior to \rmag\ as the accreting material begins to follow the field lines, the cooler of
the thermal components represents the innermost regions of the disk, just outside of
\rmag. Based on their calculations, \cite{Mushtukov17} argue that this curtain should
be optically thick, and would have the appearance of a multi-color blackbody (similar
to an accretion disk), consistent with the two thermal components typically seen in
ULX spectra. In this case, the radius of the cooler component corresponds to \rmag.
However, we note the strong similarity of the radius of the cooler thermal component
in \p13\ and the co-rotation radius (the point in the disk at which the accreting material
co-rotates with the neutron star; \rco) in this source (\citealt{Walton18p13}), which is
likely problematic for this scenario. Since accretion must be occurring in these systems
we know that \rmag\ $<$ \rco, otherwise the source would be in the `propeller' regime
in which the magnetic field prohibits accretion.

We further test the scenario proposed here by comparing the relative temperatures
and radii of the two thermal components across the full sample considered here.
Unfortunately, since the \diskbb\ component is not required in the fits for IC\,342 X-2
and Circinus ULX5 (potentially related to the high levels of absorption in these cases)
we cannot provide a reasonable comparison for these sources, but for the rest of the
sample we compute the characteristic radii of the hotter (\diskpbb/\bb, as appropriate)
and cooler (\diskbb) components similar to \cite{Walton18p13}. For the \bb\ radii, we
use standard blackbody theory, while for both the \diskbb\ and \diskpbb\ models the
normalisation is proportional to $R_{\rm{in}}^{2}\cos\theta/f_{\rm{col}}^{4}$ (where
$R_{\rm{in}}$ and $\theta$ are the inner radius and the inclination of the disk, and
$f_{\rm{col}}$ is the colour correction factor relating the observed `colour' temperature
to the effective blackbody temperature of the disk: $T_{\rm{col}} =
f_{\rm{col}}T_{\rm{eff}}$). In order to mimic our estimates for \p13, we assume that
$\cos\theta = f_{\rm{col}} = 1$ (\ie a face-on disk with no colour correction). As long as
$f_{\rm{col}}$ and $\theta$ are similar for both the inner and outer regions of the disk,
the exact values of these quantities should not strongly influence our estimates of the
\textit{relative} radii of the two components.

In all cases, we find that the radii of the hotter components are smaller than the cooler
ones, as expected. For \ngc, the radii differ by a factor of $R_{\rm{cool}}/R_{\rm{hot}}
\sim 6$, similar to \p13. However, for the remaining sample, all of which required a
\diskpbb\ component with $p < 0.75$ for the hotter component, the differences are
much larger, and are typically $R_{\rm{cool}}/R_{\rm{hot}} \sim 50$. We stress that
these comparisons are only very rough, as issues such as beaming have not been
accounted for (thus we do not compute formal errors). However, if the thick inner disk
does extend over a broader range of radii (\ie closer to the neutron star) in the ULXs
for which pulsations have not been detected, then for these sources we would also
expect the temperature of the hotter thermal component to be higher relative to that
of the cooler component. Indeed, for the known ULX pulsars we find that the ratio 
$T_{\rm{hot}}/T_{\rm{cool}} \sim 3$, while for the rest of the general ULX population
(where this can be tested) we find this ratio is $\sim$7.

To test whether these results are robust to the potential systematic issues introduced
by assuming a spectral shape for the accretion column, we investigated how the
results change for the known pulsar ULXs when the \cutoffpl\ component from the
accretion column is assumed to have the average values for $\Gamma$ and
$E_{\rm{cut}}$ used for the rest of the sample, rather than their individually
constrained spectral shapes. In addition, using Holmberg\,II X-1 as a randomly
selected test case, we also investigate how the results change when we force the
\cutoffpl\ component to have same $\Gamma$ and $E_{\rm{cut}}$ as \m82, \p13\ and
\ngc\ in turn (instead of their averages). With these tests we find that, on average,
$F_{\rm{col}}/F_{\rm{tot}}$ changes by $\sim$80\%, $R_{\rm{cool}}/R_{\rm{hot}}$
changes by $\sim$25\% and $T_{\rm{hot}}/T_{\rm{cool}}$ changes by $\sim$10\%
(where errors are given in the form ${\Delta}x/x$) relative to the results presented
above. 

Clearly the assumption made about the spectral shape of the accretion column can
have a significant effect on the relative contribution inferred for this emission. However,
we would have to be systematically underestimating $F_{\rm{col}}/F_{\rm{tot}}$ for the
unknown ULXs in order for these sources to be similar to the known ULX pulsars in this
regard. However, for the tests performed with Holmberg\,II X-1 (discussed above), we
find that when adopting the different parameter combinations from \m82, \p13\ and \ngc\
for the accretion column $F_{\rm{col}}/F_{\rm{tot}}$ both increases and decreases
(depending on the parameter combination used) in comparison to the results obtained
assuming their average spectral form. Adopting this average spectral form for the rest of
the ULX sample is therefore unlikely to systematically bias $F_{\rm{col}}/F_{\rm{tot}}$
towards lower values. Furthermore, the average changes in $R_{\rm{cool}}/R_{\rm{hot}}$
and $T_{\rm{hot}}/T_{\rm{cool}}$ are too small to change the general conclusions
discussed above.

The results presented here are therefore consistent with the idea that the main
difference between the sources where steeper and flatter radial temperature indices
are preferred for the hotter component is simply that \rsp\ and \rmag\ are closer
together in the former case, as suggested in Figure \ref{fig_schematic}, making
pulsations easier to detect. Since the former implies significant truncation by the
magnetic field, this further supports the idea that steep radial temperature indices
could also potentially indicate ULX pulsar candidates,

\subsection{ULX Demographics}

Other works have recently attempted to address the demographics of the ULX
population through theoretical considerations. \cite{Middleton17} consider the impact
of geometrical beaming on the demographics of flux-limited ULX samples within the
framework of a model in which the beaming increases with (Eddington-scaled)
accretion rate. The results are strongly dependent on the relative spatial number
densities of neutron stars and black holes, which are not currently known. However, if
this is skewed in favour of neutron stars, then scenarios in which flux-limited samples
are dominated by such sources are certainly possible. Although the sample considered
here is not formally well defined in a statistical sense, it would most likely be similar to
the flux-limited case. \cite{Wiktorowicz17} also consider this question from a stellar
evolution standpoint, performing a suite of binary population synthesis simulations.
Here, the results depend quite strongly on the star-formation history and also
somewhat on the metallicity of the host galaxies in which the ULXs reside. Continued
star formation will likely produce a ULX population dominated by neutron stars, while
brief bursts of star formation will likely produce a ULX population dominated by black
holes. In addition, lowering the metallicity increases the preference for black hole
systems. These calculations suggest that there are scenarios in which the sample
considered here could be dominated by neutron stars. Both of these theoretical
approaches are therefore consistent with the observational analysis presented here.

The calculations by \cite{Wiktorowicz17} may also offer other potential means to
identify neutron star accretors among the ULX population. Their simulations suggest
there is a channel through which the onset of a ULX phase can occur late in the
evolution of the binary system for neutron star accretors that is not available for black
holes. They predict that ULXs with evolved counterparts (red giants/supergiants;
RGs/RSGs) should only be produced by neutron star systems. A number of ULXs with
confirmed RSG counterparts have now been discovered (\citealt{Heida15, Heida16}),
with a number of additional candidates also identified (\citealt{Heida14, Lopez17}).
Neutron star accretors would provide a natural explanation for the lack of significant
variations in the radial velocities of the counterparts where multiple epochs of high S/N
spectroscopy are available (\citealt{Heida16}) as the RSG would dominate the total
mass of the system, further supporting this prediction.

\subsection{Black Hole ULXs}

Although the \nustar\ ULX sample is consistent with being dominated by neutron star
accretors, it is still possible that black hole ULXs do exist. However, to date there is
only one\footnote{The hyperluminous X-ray source ESO\,243--49 HLX1 is also widely
expected to be a black hole (being the best known candidate for an intermediate mass
black hole, with $M \sim 10^{4}$\,\msun). However, this is a clear outlier among the
broader ULX population owing to its truly extreme luminosity ($L_{\rm{X, peak}} \sim
10^{42}$\ \ergps; \citealt{Farrell09}), the fact that it exhibits X-ray spectral and
state-transitions consistent with known sub-Eddington behaviour (\citealt{Servillat11,
Davis11, Webb12}), and its (quasi-) periodic outbursts (\citealt{Godet14, Soria17}).
Instead, it is considered likely that this is the stripped nucleus of a dwarf galaxy falling
into ESO\,243--49. However, all of the methods for estimating the mass currently
available here rely on indirect scaling relations.} ULX, M101 ULX-1, in which the binary
mass function is claimed to require a black hole accretor ($M > 5$\,\msun;
\citealt{Liu13nat}). However, this is based on a very poorly-sampled radial velocity (RV)
curve. Perhaps more critically, the RV data has been compiled from \heii\ emission
assumed to arise from its Wolf-Rayet binary companion, which may lead to unreliable
mass estimates (\eg\ \citealt{Laycock15mass}). Unfortunately, this source is not
accessible to \nustar, as its spectrum is extremely soft; the source is frequently
undetected above $\sim$5\,keV (\eg\ \citealt{Soria16}). Within the framework of
high/super-Eddington accretion in which a thick, inner funnel forms in the accretion
flow, this is interpreted as the source being viewed close to edge-on, such that the
large scale-height regions of the flow obscure the hottest inner regions that would
otherwise dominate the \nustar\ band from direct view (\eg\ \citealt{Sutton13uls,
Middleton15, Pinto17}).

It is interesting to note that although all the observed spectra are qualitatively similar
at the highest energies, where the accretion column dominates in the known pulsars,
there are still differences in this band between sources in the \nustar\ sample in terms
of their long-term variability between observing epochs (typically probing
$\sim$month--year timescales). In \cite{Walton17hoIX} we found that in the case of
Holmberg\,IX X-1 the highest energies ($\gtrsim$15\,keV) remained remarkably
constant between observing epochs, despite strong variability at lower energies. In
contrast, \cite{Fuerst17ngc5907} found that the high-energy emission in \ngc\ varied
strongly in correlation with its lower energy emission. Whether this difference could be
related to different central accretors (black hole vs neutron star), or whether it could
merely be a result of, \eg, different viewing angles remains to be seen. Robust
identification of black hole ULXs that are accessible to \nustar\ will be necessary to
investigate this further.

\section{Summary and Conclusions}
\label{sec_conc}

We have undertaken a phase-resolved analysis of the ULX pulsar \ngc, following on
from our work on the other two known ULX pulsars, \m82\ and \p13. We find that the
spectral form of the pulsed emission from the accretion column is broadly similar in all
three sources (a very hard rise, before turning over to a steep spectrum at high
energies). Furthermore, we find that this emission component dominates the total
emission at the highest energies probed, resulting in the hard excesses observed in
these sources when the lower energy data are fit with accretion disk models. Extending
our consideration to the full sample of ULXs with broadband coverage to date, we find
that similar hard excesses are observed in \textit{all} the sources in this sample. For
the ULXs where the nature of the accretor is currently unknown (the majority of the
sources considered here), we investigate whether these hard excesses are all
consistent with being produced by an accretion column similar to those present in the
known ULX pulsars. We find that in all cases a similar accretion column can
successfully reproduce the observed data, using the average shape of the pulsed
emission in \m82, \p13\ and \ngc\ as a template. This is consistent with the hypothesis
that the broadband ULX sample is dominated by neutron star accretors. For the
unknown ULXs, our spectral fits suggest that the relative contribution of the non-pulsed
emission from the accretion flow beyond the magnetosphere in comparison to that
associated with the accretion columns is larger than observed in the known pulsar
ULXs. This may help to explain the lack of pulsations detected from these sources,
assuming that they are also powered by neutron star accretors.

\begin{table*}
  \caption{Details of the broadband ULX observations considered in this work}
  \vspace{-0.3cm}
\begin{center}
\begin{tabular}{c c c c c c c c}
\hline
\hline
\\[-0.1cm]
Dataset & Date & \multicolumn{3}{c}{OBSID(s)} & \multicolumn{3}{c}{Integrated Good Exposures (ks)} \\
\\[-0.2cm]
& (approx.) & \xmm\ & \suzaku\ & \nustar\ & \xmm\tmark[a] & \suzaku\ & \nustar\ \\
\\[-0.2cm]
\hline
\hline
\\[-0.1cm]
Circinus ULX5 & Feb 2013 & 0701981001 & -- & 30002038004 & 37/47 & -- & 47 \\
\\[-0.2cm]
Holmberg II X-1 & Sept 2013 & 0724810101/301 & 708015010/20 & 30001031002/3/5 & 10/23 & 90 & 262 \\
\\[-0.2cm]
Holmberg IX X-1 (low) & Apr 2015 & -- & 707019030 & 30002034006 & -- & 35 & 64 \\
\\[-0.2cm]
Holmberg IX X-1 (med) & Oct 2012 & 0693850801/0901/1001 & 707019020/30/40 & 30002033002/3 & 17/35 & 324 & 167 \\
\\[-0.2cm]
Holmberg IX X-1 (high) & Nov 2012 & 0693851101/1701/1801 & -- & 30002033005/06/08/10 & 17/24 & -- & 167 \\
\\[-0.2cm]
IC\,342 X-1 & Aug 2012 & 0693850601/1301 & -- & 30002032003/5 & 70/84 & -- & 329 \\
\\[-0.2cm]
IC\,342 X-2 & Aug 2012 & 0693850601/1301 & -- & 30002032003/5 & 70/84 & -- & 329 \\
\\[-0.2cm]
NGC\,1313 X-1 & Dec 2012 & 0693850501/1201 & -- & 30002035002/4 & 164/231 & -- & 360 \\
\\[-0.2cm]
NGC\,5204 X-1 & Apr 2013 & 0693850701/1401 & -- & 30002037002/4 & 24/30 & -- & 211 \\
\\[-0.2cm]
NGC\,5907 ULX1 (low) & Nov 2013 & 0724810401 & -- & 30002039005 & 20/32 & -- & 123 \\
\\[-0.2cm]
NGC\,5907 ULX1 (high) & July 2014 & 0729561301 & -- & 80001042002/4 & 38/43 & -- & 132 \\
\\[-0.2cm]
NGC\,7793 P13 & May 2016 & 0781800101 & -- & 80201010002 & 28/46 & -- & 114 \\
\\[-0.2cm]
\hline
\hline
\\[-0.35cm]
\end{tabular}
\label{tab_obslog}
\end{center}
$^{a}$ \xmm\ exposures are listed for the \epicpn/MOS detectors 
\vspace{0.3cm}
\end{table*}

\section*{ACKNOWLEDGEMENTS}

The authors would like to thank the reviewer for their positive feedback, which helped
to improve the final manuscript. DJW and MJM acknowledge support from STFC
Ernest Rutherford fellowships, ACF acknowledges support from ERC Advanced Grant
340442, and DB acknowledges financial support from the French Space Agency
(CNES). This research has made use of data obtained with \nustar, a project led by
Caltech, funded by NASA and managed by NASA/JPL, and has utilized the \nustardas\
software package, jointly developed by the ASDC (Italy) and Caltech (USA). This work
has also made use of data obtained with \xmm, an ESA science mission with
instruments and contributions directly funded by ESA Member States, and with \suzaku,
a collaborative mission between the space agencies of Japan (JAXA) and the USA
(NASA).

{\it Facilites:} \facility{NuSTAR}, \facility{XMM}, \facility{Suzaku}

\appendix

\section{A. Broadband Observation Log}
\label{app_obslog}

Table \ref{tab_obslog} gives the key details for the \xmm, \nustar\ and \suzaku\
observations that comprise the broadband ULX datasets considered in this work. \\

\section{B. \textit{NuSTAR} Pulsation Search}
\label{app_pulse}

The \xmm\ data considered in this work has been systematically searched for
pulsations by either \cite{Doroshenko15} and/or the ExTRAS variability
survey\footnote{http://www.extras-fp7.eu/index.php} (\citealt{EXTRAS}). Aside from
the known pulsar systems, no further coherent signals have been significantly
detected in the rest of the broadband ULX sample to date. We have also extended
these searches into the \nustar\ data considered here, following the approach taken
in \cite{Fuerst16p13} for \p13. For each \nustar\ dataset we calculated power
spectral densities (PSDs) based on 3--40\,keV light curves with 0.05\,s resolution; all
times were transferred to the solar barycenter using the DE200 solar ephemeris. We
searched the complete available frequency range up to a maximum frequency of
10\,Hz using the \accelsearch\ routine in \presto\ (\citealt{PRESTOaccel}), following
the approach pioneered in \cite{Bachetti14nat}. Aside from the known pulsar systems,
no significant excess over the Poisson noise was found in any observation for any
other source considered here, even when allowing for a period derivative in these
searches. This is consistent with the results from \xmm.

\bibliographystyle{/Users/dwalton/papers/mnras}

\bibliography{/Users/dwalton/papers/references}

\begin{thebibliography}{74}
\expandafter\ifx\csname natexlab\endcsname\relax\def\natexlab#1{#1}\fi

\bibitem[{Abramowicz} et~al.(1988){Abramowicz}, {Czerny}, {Lasota} \&
  {Szuszkiewicz}]{Abram88}
{Abramowicz} M.~A., {Czerny} B., {Lasota} J.~P., {Szuszkiewicz} E., 1988, \apj,
  332, 646

\bibitem[{Bachetti} et~al.(2014){Bachetti}, {Harrison}, {Walton}
  et~al.]{Bachetti14nat}
{Bachetti} M., {Harrison} F.~A., {Walton} D.~J., et~al., 2014, \nat, 514, 202

\bibitem[{Bachetti} et~al.(2013){Bachetti}, {Rana}, {Walton}
  et~al.]{Bachetti13}
{Bachetti} M., {Rana} V., {Walton} D.~J., et~al., 2013, \apj, 778, 163

\bibitem[{Brightman} et~al.(2016{\natexlab{a}}){Brightman}, {Harrison},
  {Walton} et~al.]{Brightman16m82a}
{Brightman} M., {Harrison} F., {Walton} D.~J., et~al., 2016{\natexlab{a}},
  \apj, 816, 60

\bibitem[{Brightman} et~al.(2016{\natexlab{b}}){Brightman}, {Harrison},
  {Barret} et~al.]{Brightman16m82b}
{Brightman} M., {Harrison} F.~A., {Barret} D., et~al., 2016{\natexlab{b}},
  \apj, 829, 28

\bibitem[{Cui}(1997)]{Cui97}
{Cui} W., 1997, \apjl, 482, L163

\bibitem[{Davis} et~al.(2011){Davis}, {Narayan}, {Zhu} et~al.]{Davis11}
{Davis} S.~W., {Narayan} R., {Zhu} Y., et~al., 2011, \apj, 734, 111

\bibitem[{De Luca} et~al.(2016){De Luca}, {Salvaterra}, {Tiengo}
  et~al.]{EXTRAS}
{De Luca} A., {Salvaterra} R., {Tiengo} A., et~al., 2016, The Universe of
  Digital Sky Surveys, 42, 291

\bibitem[{Doroshenko} et~al.(2015){Doroshenko}, {Santangelo} \&
  {Ducci}]{Doroshenko15}
{Doroshenko} V., {Santangelo} A., {Ducci} L., 2015, \aap, 579, A22

\bibitem[{Dotan} \& {Shaviv}(2011)]{Dotan11}
{Dotan} C., {Shaviv} N.~J., 2011, \mnras, 413, 1623

\bibitem[{Farrell} et~al.(2009){Farrell}, {Webb}, {Barret}, {Godet} \&
  {Rodrigues}]{Farrell09}
{Farrell} S.~A., {Webb} N.~A., {Barret} D., {Godet} O., {Rodrigues} J.~M.,
  2009, \nat, 460, 73

\bibitem[{Freeman} et~al.(1977){Freeman}, {Karlsson}, {Lynga}
  et~al.]{Freeman77}
{Freeman} K.~C., {Karlsson} B., {Lynga} G., et~al., 1977, \aap, 55, 445

\bibitem[{F{\"u}rst} et~al.(2016){F{\"u}rst}, {Walton}, {Harrison}
  et~al.]{Fuerst16p13}
{F{\"u}rst} F., {Walton} D.~J., {Harrison} F.~A., et~al., 2016, \apjl, 831, L14

\bibitem[{F{\"u}rst} et~al.(2017){F{\"u}rst}, {Walton}, {Stern}
  et~al.]{Fuerst17ngc5907}
{F{\"u}rst} F., {Walton} D.~J., {Stern} D., et~al., 2017, \apj, 834, 77

\bibitem[{Gladstone} et~al.(2013){Gladstone}, {Copperwheat}, {Heinke}
  et~al.]{Gladstone13}
{Gladstone} J.~C., {Copperwheat} C., {Heinke} C.~O., et~al., 2013, \apjs, 206,
  14

\bibitem[{Godet} et~al.(2014){Godet}, {Lombardi}, {Antonini} et~al.]{Godet14}
{Godet} O., {Lombardi} J.~C., {Antonini} F., et~al., 2014, \apj, 793, 105

\bibitem[{Harrison} et~al.(2013){Harrison}, {Craig}, {Christensen}
  et~al.]{NUSTAR}
{Harrison} F.~A., {Craig} W.~W., {Christensen} F.~E., et~al., 2013, \apj, 770,
  103

\bibitem[{Heida} et~al.(2014){Heida}, {Jonker}, {Torres} et~al.]{Heida14}
{Heida} M., {Jonker} P.~G., {Torres} M.~A.~P., et~al., 2014, \mnras, 442, 1054

\bibitem[{Heida} et~al.(2016){Heida}, {Jonker}, {Torres} et~al.]{Heida16}
{Heida} M., {Jonker} P.~G., {Torres} M.~A.~P., et~al., 2016, \mnras, 459, 771

\bibitem[{Heida} et~al.(2015){Heida}, {Torres}, {Jonker} et~al.]{Heida15}
{Heida} M., {Torres} M.~A.~P., {Jonker} P.~G., et~al., 2015, \mnras, 453, 3511

\bibitem[{Israel} et~al.(2017{\natexlab{a}}){Israel}, {Belfiore}, {Stella}
  et~al.]{Israel17}
{Israel} G.~L., {Belfiore} A., {Stella} L., et~al., 2017{\natexlab{a}},
  Science, 355, 817

\bibitem[{Israel} et~al.(2017{\natexlab{b}}){Israel}, {Papitto}, {Esposito}
  et~al.]{Israel17p13}
{Israel} G.~L., {Papitto} A., {Esposito} P., et~al., 2017{\natexlab{b}},
  \mnras, 466, L48

\bibitem[{Jansen} et~al.(2001){Jansen}, {Lumb}, {Altieri} et~al.]{XMM}
{Jansen} F., {Lumb} D., {Altieri} B., et~al., 2001, \aap, 365, L1

\bibitem[{Kaaret} et~al.(2009){Kaaret}, {Feng} \& {Gorski}]{Kaaret09m82}
{Kaaret} P., {Feng} H., {Gorski} M., 2009, \apj, 692, 653

\bibitem[{Kalberla} et~al.(2005){Kalberla}, {Burton}, {Hartmann} et~al.]{NH}
{Kalberla} P.~M.~W., {Burton} W.~B., {Hartmann} D., et~al., 2005, \aap, 440,
  775

\bibitem[{Karachentsev} et~al.(2002){Karachentsev}, {Dolphin}, {Geisler}
  et~al.]{Karachentsev02}
{Karachentsev} I.~D., {Dolphin} A.~E., {Geisler} D., et~al., 2002, \aap, 383,
  125

\bibitem[{King} et~al.(2017){King}, {Lasota} \& {Klu{\'z}niak}]{King17ulx}
{King} A., {Lasota} J.-P., {Klu{\'z}niak} W., 2017, \mnras, 468, L59

\bibitem[{Koliopanos} et~al.(2017){Koliopanos}, {Vasilopoulos}, {Godet},
  {Bachetti}, {Webb} \& {Barret}]{Koliopanos17}
{Koliopanos} F., {Vasilopoulos} G., {Godet} O., {Bachetti} M., {Webb} N.~A.,
  {Barret} D., 2017, \aap, 608, A47

\bibitem[{Lamb} et~al.(1973){Lamb}, {Pethick} \& {Pines}]{Lamb73}
{Lamb} F.~K., {Pethick} C.~J., {Pines} D., 1973, \apj, 184, 271

\bibitem[{Laycock} et~al.(2015){Laycock}, {Maccarone} \&
  {Christodoulou}]{Laycock15mass}
{Laycock} S.~G.~T., {Maccarone} T.~J., {Christodoulou} D.~M., 2015, \mnras,
  452, L31

\bibitem[{Liu} et~al.(2013){Liu}, {Bregman}, {Bai}, {Justham} \&
  {Crowther}]{Liu13nat}
{Liu} J.-F., {Bregman} J.~N., {Bai} Y., {Justham} S., {Crowther} P., 2013,
  \nat, 503, 500

\bibitem[{L{\'o}pez} et~al.(2017){L{\'o}pez}, {Heida}, {Jonker}
  et~al.]{Lopez17}
{L{\'o}pez} K.~M., {Heida} M., {Jonker} P.~G., et~al., 2017, ArXiv 1704.01068

\bibitem[{Luangtip} et~al.(2016){Luangtip}, {Roberts} \& {Done}]{Luangtip16}
{Luangtip} W., {Roberts} T.~P., {Done} C., 2016, \mnras

\bibitem[{M{\'e}ndez} et~al.(2002){M{\'e}ndez}, {Davis}, {Moustakas}, {Newman},
  {Madore} \& {Freedman}]{Mendez02}
{M{\'e}ndez} B., {Davis} M., {Moustakas} J., {Newman} J., {Madore} B.~F.,
  {Freedman} W.~L., 2002, \aj, 124, 213

\bibitem[{Middleton} \& {King}(2017)]{Middleton17}
{Middleton} M., {King} A., 2017, ArXiv e-prints

\bibitem[{Middleton} et~al.(2015){Middleton}, {Heil}, {Pintore}, {Walton} \&
  {Roberts}]{Middleton15}
{Middleton} M.~J., {Heil} L., {Pintore} F., {Walton} D.~J., {Roberts} T.~P.,
  2015, \mnras, 447, 3243

\bibitem[{Miller} et~al.(2014){Miller}, {Bachetti}, {Barret} et~al.]{Miller14}
{Miller} J.~M., {Bachetti} M., {Barret} D., et~al., 2014, \apjl, 785, L7

\bibitem[{Miller} et~al.(2013){Miller}, {Walton}, {King} et~al.]{Miller13ulx}
{Miller} J.~M., {Walton} D.~J., {King} A.~L., et~al., 2013, \apjl, 776, L36

\bibitem[{Mitsuda} et~al.(2007){Mitsuda}, {Bautz}, {Inoue} et~al.]{SUZAKU}
{Mitsuda} K., {Bautz} M., {Inoue} H., et~al., 2007, \pasj, 59, 1

\bibitem[{Motch} et~al.(2014){Motch}, {Pakull}, {Soria}, {Gris{\'e}} \&
  {Pietrzy{\'n}ski}]{Motch14nat}
{Motch} C., {Pakull} M.~W., {Soria} R., {Gris{\'e}} F., {Pietrzy{\'n}ski} G.,
  2014, \nat, 514, 198

\bibitem[{Mukherjee} et~al.(2015){Mukherjee}, {Walton}, {Bachetti}
  et~al.]{Mukherjee15}
{Mukherjee} E.~S., {Walton} D.~J., {Bachetti} M., et~al., 2015, \apj, 808, 64

\bibitem[{Mushtukov} et~al.(2017){Mushtukov}, {Suleimanov}, {Tsygankov} \&
  {Ingram}]{Mushtukov17}
{Mushtukov} A.~A., {Suleimanov} V.~F., {Tsygankov} S.~S., {Ingram} A., 2017,
  \mnras, 467, 1202

\bibitem[{Paturel} et~al.(2002){Paturel}, {Theureau}, {Fouqu{\'e}}, {Terry},
  {Musella} \& {Ekholm}]{Paturel02}
{Paturel} G., {Theureau} G., {Fouqu{\'e}} P., {Terry} J.~N., {Musella} I.,
  {Ekholm} T., 2002, \aap, 383, 398

\bibitem[{Pietrzy{\'n}ski} et~al.(2010){Pietrzy{\'n}ski}, {Gieren}, {Hamuy}
  et~al.]{Pietrzynski10}
{Pietrzy{\'n}ski} G., {Gieren} W., {Hamuy} M., et~al., 2010, \aj, 140, 1475

\bibitem[{Pinto} et~al.(2017){Pinto}, {Alston}, {Soria} et~al.]{Pinto17}
{Pinto} C., {Alston} W., {Soria} R., et~al., 2017, \mnras, 468, 2865

\bibitem[{Pinto} et~al.(2016){Pinto}, {Middleton} \& {Fabian}]{Pinto16nat}
{Pinto} C., {Middleton} M.~J., {Fabian} A.~C., 2016, \nat, 533, 64

\bibitem[{Pintore} et~al.(2017){Pintore}, {Zampieri}, {Stella}, {Wolter},
  {Mereghetti} \& {Israel}]{Pintore17}
{Pintore} F., {Zampieri} L., {Stella} L., {Wolter} A., {Mereghetti} S.,
  {Israel} G.~L., 2017, ArXiv e-prints

\bibitem[{Poutanen} et~al.(2007){Poutanen}, {Lipunova}, {Fabrika}, {Butkevich}
  \& {Abolmasov}]{Poutanen07}
{Poutanen} J., {Lipunova} G., {Fabrika} S., {Butkevich} A.~G., {Abolmasov} P.,
  2007, \mnras, 377, 1187

\bibitem[{Rana} et~al.(2015){Rana}, {Harrison}, {Bachetti} et~al.]{Rana15}
{Rana} V., {Harrison} F.~A., {Bachetti} M., et~al., 2015, \apj, 799, 121

\bibitem[{Ransom} et~al.(2002){Ransom}, {Eikenberry} \&
  {Middleditch}]{PRESTOaccel}
{Ransom} S.~M., {Eikenberry} S.~S., {Middleditch} J., 2002, \aj, 124, 1788

\bibitem[{Saha} et~al.(2002){Saha}, {Claver} \& {Hoessel}]{Saha02}
{Saha} A., {Claver} J., {Hoessel} J.~G., 2002, \aj, 124, 839

\bibitem[{Servillat} et~al.(2011){Servillat}, {Farrell}, {Lin}, {Godet},
  {Barret} \& {Webb}]{Servillat11}
{Servillat} M., {Farrell} S.~A., {Lin} D., {Godet} O., {Barret} D., {Webb}
  N.~A., 2011, \apj, 743, 6

\bibitem[{Shakura} \& {Sunyaev}(1973)]{Shakura73}
{Shakura} N.~I., {Sunyaev} R.~A., 1973, \aap, 24, 337

\bibitem[{Soria} \& {Kong}(2016)]{Soria16}
{Soria} R., {Kong} A., 2016, \mnras, 456, 1837

\bibitem[{Soria} et~al.(2017){Soria}, {Musaeva}, {Wu} et~al.]{Soria17}
{Soria} R., {Musaeva} A., {Wu} K., et~al., 2017, \mnras, 469, 886

\bibitem[{Steiner} et~al.(2009){Steiner}, {Narayan}, {McClintock} \&
  {Ebisawa}]{SIMPL}
{Steiner} J.~F., {Narayan} R., {McClintock} J.~E., {Ebisawa} K., 2009, \pasp,
  121, 1279

\bibitem[{Str{\"u}der} et~al.(2001){Str{\"u}der}, {Briel}, {Dennerl}
  et~al.]{XMM_PN}
{Str{\"u}der} L., {Briel} U., {Dennerl} K., et~al., 2001, \aap, 365, L18

\bibitem[{Sutton} et~al.(2013{\natexlab{a}}){Sutton}, {Roberts}, {Gladstone}
  et~al.]{Sutton13}
{Sutton} A.~D., {Roberts} T.~P., {Gladstone} J.~C., et~al., 2013{\natexlab{a}},
  \mnras, 434, 1702

\bibitem[{Sutton} et~al.(2013{\natexlab{b}}){Sutton}, {Roberts} \&
  {Middleton}]{Sutton13uls}
{Sutton} A.~D., {Roberts} T.~P., {Middleton} M.~J., 2013{\natexlab{b}}, \mnras,
  435, 1758

\bibitem[{Sutton} et~al.(2012){Sutton}, {Roberts}, {Walton}, {Gladstone} \&
  {Scott}]{Sutton12}
{Sutton} A.~D., {Roberts} T.~P., {Walton} D.~J., {Gladstone} J.~C., {Scott}
  A.~E., 2012, \mnras, 423, 1154

\bibitem[{Swartz} et~al.(2004){Swartz}, {Ghosh}, {Tennant} \& {Wu}]{Swartz04}
{Swartz} D.~A., {Ghosh} K.~K., {Tennant} A.~F., {Wu} K., 2004, \apjs, 154, 519

\bibitem[{Tully} et~al.(2016){Tully}, {Courtois} \& {Sorce}]{Tully16}
{Tully} R.~B., {Courtois} H.~M., {Sorce} J.~G., 2016, \aj, 152, 50

\bibitem[{Verner} et~al.(1996){Verner}, {Ferland}, {Korista} \&
  {Yakovlev}]{Verner96}
{Verner} D.~A., {Ferland} G.~J., {Korista} K.~T., {Yakovlev} D.~G., 1996, \apj,
  465, 487

\bibitem[{Walton} et~al.(2013){Walton}, {Fuerst}, {Harrison}
  et~al.]{Walton13culx}
{Walton} D.~J., {Fuerst} F., {Harrison} F., et~al., 2013, \apj, 779, 148

\bibitem[{Walton} et~al.(2017){Walton}, {F{\"u}rst}, {Harrison}
  et~al.]{Walton17hoIX}
{Walton} D.~J., {F{\"u}rst} F., {Harrison} F.~A., et~al., 2017, \apj, 839, 105

\bibitem[{Walton} et~al.(2018){Walton}, {F{\"u}rst}, {Harrison}
  et~al.]{Walton18p13}
{Walton} D.~J., {F{\"u}rst} F., {Harrison} F.~A., et~al., 2018, \mnras, 473,
  4360

\bibitem[{Walton} et~al.(2015{\natexlab{a}}){Walton}, {Harrison}, {Bachetti}
  et~al.]{Walton15}
{Walton} D.~J., {Harrison} F.~A., {Bachetti} M., et~al., 2015{\natexlab{a}},
  \apj, 799, 122

\bibitem[{Walton} et~al.(2014){Walton}, {Harrison}, {Grefenstette}
  et~al.]{Walton14hoIX}
{Walton} D.~J., {Harrison} F.~A., {Grefenstette} B.~W., et~al., 2014, \apj,
  793, 21

\bibitem[{Walton} et~al.(2016){Walton}, {Middleton}, {Pinto}
  et~al.]{Walton16ufo}
{Walton} D.~J., {Middleton} M.~J., {Pinto} C., et~al., 2016, \apjl, 826, L26

\bibitem[{Walton} et~al.(2015{\natexlab{b}}){Walton}, {Middleton}, {Rana}
  et~al.]{Walton15hoII}
{Walton} D.~J., {Middleton} M.~J., {Rana} V., et~al., 2015{\natexlab{b}}, \apj,
  806, 65

\bibitem[{Walton} et~al.(2011){Walton}, {Roberts}, {Mateos} \&
  {Heard}]{WaltonULXCat}
{Walton} D.~J., {Roberts} T.~P., {Mateos} S., {Heard} V., 2011, \mnras, 416,
  1844

\bibitem[{Webb} et~al.(2012){Webb}, {Cseh}, {Lenc} et~al.]{Webb12}
{Webb} N., {Cseh} D., {Lenc} E., et~al., 2012, Science, 337, 554

\bibitem[{Wiktorowicz} et~al.(2017){Wiktorowicz}, {Sobolewska}, {Lasota} \&
  {Belczynski}]{Wiktorowicz17}
{Wiktorowicz} G., {Sobolewska} M., {Lasota} J.-P., {Belczynski} K., 2017, ArXiv
  e-prints

\bibitem[{Wilms} et~al.(2000){Wilms}, {Allen} \& {McCray}]{tbabs}
{Wilms} J., {Allen} A., {McCray} R., 2000, \apj, 542, 914

\end{thebibliography}

\label{lastpage}

\end{document}